\documentclass[20pt-article, superscriptaddress,showpacs,nofootinbib,floatfix,
preprintnumbers]{revtex4-1}
\usepackage{epsfig,amsfonts,amsthm,amssymb}
\usepackage{booktabs,color}
\newcommand{\be}{\begin{equation}}
\newcommand{\ee}{\end{equation}}
\newcommand{\bea}{\begin{eqnarray}}
\newcommand{\eea}{\end{eqnarray}}

\def\lsim{\mathrel{\rlap{\lower4pt\hbox{\hskip1pt$\sim$}}
    \raise1pt\hbox{$<$}}}         %less than or approx. symbol
\def\gsim{\mathrel{\rlap{\lower4pt\hbox{\hskip1pt$\sim$}}
    \raise1pt\hbox{$>$}}}         %greater than or approx. symbol

\begin{document}

\thispagestyle{empty}
\begin{flushright}
\texttt{IPPP/13/90}\\
\texttt{DCPT/13/180}\\
\end{flushright}
\bigskip

\begin{center}
{\Large{\bf The Effect of Cancellation in Neutrinoless Double Beta Decay}}

\vskip 0.5cm

{Silvia Pascoli$^{a}$, 
Manimala Mitra$^a$,
 Steven Wong $^{b}$ 
}
\vskip .5cm
$^a${ Institute for Particle Physics Phenomenology, 
Department of Physics, \\ Durham University, Durham DH1 3LE, United Kingdom}

$^b${ Physics Department, Chinese University of Hong Kong, Shatin, N.T., Hong Kong}

\vskip 0.7cm

{\bf Abstract}
\end{center}
\vskip 0.3cm

In  light of  recent experimental results, we carefully analyze the effects  of interference  in neutrinoless double beta decay,
 when   more than one mechanism is operative.  
If a complete cancellation is at work, the half-life of the corresponding isotope is infinite and any constraint on it
will automatically be satisfied. We analyze this possibility in detail assuming  a cancellation in $^{136}\rm{Xe}$, and 
find its   implications  on the half-life of 
 other isotopes, such as $^{76}\rm{Ge}$. 
For definiteness, we consider  the role  of light and heavy sterile neutrinos. 
In this case, the effective Majorana mass parameter can be  redefined to 
take into account all contributions and 
 its value gets suppressed. 
Hence, larger values of neutrino masses are required  for  the same half-life. 
The canonical light neutrino contribution can not
saturate the present limits of half-lives or the positive claim 
 of observation of neutrinoless double beta decay, once the stringent bounds from cosmology 
are  taken into account. 
For the case of cancellation, where all the sterile neutrinos are heavy,  
the tension between the results  from neutrinoless double beta decay and   cosmology 
 becomes  more severe.  We show that the  inclusion of  light sterile neutrinos 
in this set up can resolve this  issue. Using the recent results from GERDA,  we derive upper limits 
 on the  active-sterile 
mixing angles and compare it with the case of no cancellation. 
The required values of the mixing angles 
become  larger, if a cancellation is at work. 
A direct test of destructive interference in $^{136}\rm{Xe}$  is provided by the 
observation of this process in other isotopes and we study in detail  the correlation between   
their half-lives. 
Finally, we discuss the  model realizations  which can accommodate light and heavy sterile neutrinos 
and the cancellation.  We show that sterile neutrinos of few hundred MeV or GeV mass range,  coming  from  an 
 Extended seesaw framework or a  further extension,  can satisfy the 
required cancellation.

\maketitle

%%%%%%%%%%%%%%%%%%%%%%%%%%%%%%
\section{Introduction \label{intro}} 

In the past fifteen years, the experimental evidence of neutrino masses and mixing has opened up a new window on the physics beyond the standard model. The  solar, atmospheric and reactor  neutrino  oscillation 
(see \cite{review-osc-petcov, mik-shap,  gelmini-rev,  nir-garcia-rev,  rev-moha-smir, visreview} for recent reviews) experiments \cite{solar,kl, atm, k2k, t2k, minos,chooz} of the past decades confirmed that the standard neutrinos have very small  masses in the eV range. Neutrino mixing data \cite{limits, limitspres, Fogli2011osc, oscparam}
 is well described by the unitary PMNS matrix $U$,  parameterized
by three real mixing angles, one CP violating  Dirac phase  and two Majorana phases. So far, the  oscillation parameters, namely
the solar, atmospheric mass square differences 
$\Delta m^2_{12}$ and $\Delta m^2_{13}$ and the three oscillation angles $\theta_{12}$, $\theta_{13}$ and $\theta_{23}$, have 
been measured \cite{solar, atm, kl,  k2k,t2k,minos, chooz, limits, limitspres, Fogli2011osc, oscparam,  reno, dayabay, dc} upto a good accuracy.  The current $3 \sigma$ allowed ranges of the oscillation parameters are  \cite{limitspres, oscparam}
%%%%%%
\be
 6.99 \times 10^{-5} \rm{eV^2} \leq \Delta m_{21}^2 \leq 8.18\times 10^{-5} \rm{eV^2},\hspace*{0.5cm} 2.17 \times 10^{-3} \rm{eV^2} \leq \Delta m_{31}^2 
\leq 2.62\times 10^{-3}  \rm{eV^2} ~, \ee 
%%%%%%
%%%%%%
\be
   0.259 \leq  \sin^2 \theta_{12} \leq 0.359,\hspace*{0.5cm}    0.331 \leq \sin^2 \theta_{23} \leq 0.663 , \hspace*{0.5cm}  0.016\leq  \sin^2 \theta_{13} \leq 0.031 ~.
\ee 
%%%%%%
%\end{document}
Although a lot of information on neutrino masses and mixing have been unveiled in the past  decade, yet many neutrino properties remain to be determined. We still 
do not know the neutrino mass hierarchy, if the CP symmetry is violated in the leptonic sector, and most importantly, the nature of neutrinos - whether neutrinos are 
Dirac or Majorana particles. 
The neutrino nature is strictly related to the violation of global leptonic number and, hence,  experiments in which lepton number violation can manifest itself could unveil the Majorana nature of neutrinos. 

Among the different lepton number violating experiments, neutrinoless double beta decay, searching for $(A,Z) \to (A,Z+2)+2e^-$, \cite{oli-rev, vogel-rev, bilenky-rev, werner-rev, klapdorold, klapdor, Gda, gerda, hei, igex, Gando:2012zm, Auger:2012ar, cuo, nemo3} is  the 
most sensitive one. In the minimal extension of the Standard Model, augmented only by massive neutrinos, this process is mediated by light neutrino exchange \cite{0nu2beta-old}. In this case
the observation of $(\beta \beta)_{0\nu}$-decay can shed some light on i) the mass hierarchy, the neutrino mass scale and, possibly, on one of the Majorana CP-violating phases, although this will be very challenging \cite{silvia0nu2beta, Barger:2002vy}. However, in general other mechanisms could play a role in neutrinoless double beta decay. In fact,
Majorana neutrino masses require further extensions of the standard model, with a new physics scale, new particles and a source of lepton number violation. The simplest realization comes from the dimension 5 operator $L \cdot H L\cdot H / \Lambda$ \cite{dim5}, which can arise as the low energy effective term from a higher energy theory with lepton number violation. The latter will typically also induce neutrinoless double beta decay directly. In most cases, such contributions are suppressed due to the heavy scale of the new mediators, but many exceptions exist \cite{feinberg}. 
Several  detailed studies have been carried out \cite{f2005, ibarra, MSV, msvmoriond, Pascoli,  blennow, meroni2013} 
 regarding  Type-I \cite{seesawM, seesaw-Goran, seesaw-Yan,seesaw-Ramond}, Extended \cite{Kang-Kim, Parida} and Inverse seesaw \cite{invo, inv, invdet,  inverseso10, inverseothers}, 
Left-Right symmetric \cite{LRSM, Tello:2010,Nemevsek:2011aa, Chakrabortty:2012mh, Barry:2013xxa,Dev:2013vxa,Huang:2013kma,Dev:2013oxa}, R-parity violating supersymmetric models \cite{ms0nu2beta, mwex, pion-ex, hirsch,  allanach}.
It is found that in the Type-I  and  Extended seesaw scenario
sterile neutrinos with few GeV masses can give a contribution comparable to the light neutrino ones or even dominant~\cite{MSV,msvmoriond,Pascoli}. 
For Left-Right symmetric models, the right handed current 
contribution can be significantly large,  if the new gauge boson and right handed neutrino 
masses are in the TeV scale \cite{Tello:2010,Nemevsek:2011aa,Chakrabortty:2012mh, Barry:2013xxa,Dev:2013vxa,Huang:2013kma,Dev:2013oxa}. 
In the case of R-parity violating supersymmetry, different lepton number violating states e.g. neutralino, squark and gluino can mediate this process, 
and their contributions have been analyzed in detail 
\cite{ms0nu2beta, mwex, pion-ex, hirsch,  allanach}.
The different lepton number violating states can also originate from an extra dimensional framework \cite{extra} or other possible new physics scenario \cite{vogel, choi, Choubey:2012ux}.

Several experiments on neutrinoless double beta decay 
\cite{klapdorold, klapdor, Gda, gerda, hei, igex, Gando:2012zm,Auger:2012ar, cuo, nemo3} have been carried out using different type of nuclei, e.g. $^{76}\rm{Ge}$, $^{136}\rm{Xe}$, $^{100}\rm{Mo}$, $^{130}\rm{Te}$.
The  bounds coming  from Heidelberg-Moscow \cite{hei} and IGEX \cite{igex}
 experiments apply to the $^{76}\rm{Ge}$ isotope and are given by 
$T^{0\nu}_{1/2}> (1.9, 1.55) \times 10^{25} \rm{yrs}$ at $90\% $C.L., respectively, but the most stringent bound  has been recently reported by the GERDA collaboration: $T^{0\nu}_{1/2}> 2.1\times 10^{25} \rm{yrs}$ at 
$90\%$ C.L. \cite{gerda}. Combining the latter with the Heidelberg-Moscow and IGEX experiments, the limit improves to $T^{0\nu}_{1/2}> 3.0\times 10^{25} \rm{yrs}$ at 
$90\%$ C.L. \cite{gerda}. It should be pointed out that a part of the Heidelberg-Moscow
collaboration, led by Klapdor-Kleingrothaus and collaborators, reported evidence of 
the observation of this process  corresponding  to the half-life $T^{0\nu}_{1/2}(^{76}\rm{Ge})=1.19^{+0.37}_{-0.23} \times 10^{25} \rm{yrs}$ \cite{klapdorold}, which   was updated later to $T^{0\nu}_{1/2}(^{76}\rm{Ge})=2.23^{+0.44}_{-0.31} \times 10^{25} \rm{yrs}$ \cite{klapdor}. This claim 
 has been constrained significantly by the recent results from GERDA \cite{gerda} 
but at present neither the individual nor the combined limit from
GERDA [24] can conclusively rule  out the updated claim \cite{klapdor}.
Using the $^{136}\rm{Xe}$ isotope,
 the bounds on half-life from EXO-200 and KamLAND-Zen experiments are 
$T^{0\nu}_{1/2}> 1.6\times 10^{25} \rm{yrs}$ \cite{Auger:2012ar}  and $T^{0\nu}_{1/2}> 1.9\times 10^{25} \rm{yrs}$ \cite{Gando:2012zm} at $90\%$ C.L., respectively. The KamLAND-Zen collaboration has combined the 
 two limits  obtaining  $T^{0\nu}_{1/2}> 3.4\times 10^{25} \rm{yrs}$ at $90\%$ C.L. \cite{Gando:2012zm}.
According to the KamLAND-Zen collaboration this combined bound rules out the claim in \cite{klapdor} at $99.7\%$ C.L. but, as pointed out in \cite{Dev:2013vxa}, this conclusion 
depends  on the nuclear matrix elements (NME) used. Future experiments will conclusively confirm or disprove the positive claim and can improve the sensitivity 
to the half-life by more than an order of magnitude \cite{Gda,  Cuore,  Gando:2012zm, supernemo, Majorana0, lucifer, snop, cobra, next}.

The light neutrinos, if Majorana particle, will mediate the neutrinoless double beta decay. Their contribution can saturate the present limits of half-lives only in the quasi-degenerate limit.
As pointed out in Ref.~\cite{msvmoriond,Dev:2013vxa,fogli}, the bounds from cosmology put stringent constraint on neutrino masses and consequently on the interpretation of neutrinoless double beta decay mediated by light neutrino masses to satisfy the claim in \cite{klapdor}, or to saturate
the experimental limits from Heidelberg-Moscow, GERDA, EXO-200 and  KamLAND-Zen \cite{gerda, hei, Gando:2012zm, Auger:2012ar}. The conclusion remains 
the same, after including the stringent cosmological bound on the sum of light neutrino masses from Planck 
\cite{Ade:2013lta}, as it has been explicitly shown in \cite{Dev:2013vxa}.

In the light of the recent experimental results, in this work  we carefully analyze  lepton number violation in neutrinoless double beta decay for the cases in which more than one mechanism is operative~\cite{Pascoli}. {In presence of several left-current processes, if their contributions are comparable, they can sum up constructively in neutrinoless double beta decay or even partially or completely cancel out, making the half-life much longer than naively expected. Establishing if cancellations are at play could be of importance to conclusively determine the nature of neutrinos. In fact, if future experiments do not find neutrinoless double beta decay in contradiction with the theoretical prediction, 
the conclusion that neutrinos are Dirac particles is valid only if 
the possibility of cancellations between different mechanisms is excluded. For instance, this would be the case if no positive evidence is found down to an effective Majorana mass parameter of 10 meV and an inverted hierarchy is established in reactor, atmospheric and/or long baseline neutrino oscillation experiments. 
Here, we show that if both light and heavy neutrinos, compared to the momentum exchange of the process, are at work, it might be possible to test the presence of such a cancellation.

While individual contribution from different underlying mechanisms: e.g. the most popular light neutrinos, sterile neutrinos in Type-I, Extended seesaw and Inverse seesaw, gluino and squark exchange for R-parity violating supersymmetry, have been carefully analyzed in the literature, the interference effects have been neglected to a large extent  (see  \cite{petcov, Faessler:2011qw, Faessler:2011rv, f2010} for the few  discussions on the interference). In this work, we discuss the effect of interference  in detail and present simple model realizations in which such cancellations can emerge. Although our analysis is general, one immediate application would be to solve the mutual inconsistency between 
the  positive claim in \cite{klapdor} for $^{76}\rm{Ge}$ and the bounds from \cite{Auger:2012ar,Gando:2012zm} in $^{136}\rm{Xe}$. If the found evidence \cite{klapdor} is finally refuted by future experiments, the possibility of cancellations remains open and should be tested by using different nuclei.

The paper is organized as follows. In Section \ref{lightneu}, we review the different bounds on neutrinoless double beta decay; we discuss the contribution from light neutrino exchange,  the stringent bounds on neutrino masses from cosmology as well as the  future bound from KATRIN \cite{katrin}. Following that, we discuss the contribution from sterile neutrinos  in  Section \ref{largesterile}. 
We discuss  the cancellations in Section \ref{cance}, where we carefully consider the interference between  two dominant mechanisms in neutrinoless double beta decay, e.g. light neutrino-heavy sterile neutrino exchange or light neutrino-gluino/squark exchange. We show how this possibility is further constrained from beta decay as well as cosmology. Next, we consider the case in which both light and heavy sterile neutrinos are operative in neutrinoless double beta decay. This possibility allows to overcome the  constraints from  cosmology.  We discuss the correlation of half-lives between two different isotopes in Section \ref{cor}. In Section \ref{mod} we discuss simple model realizations which can accommodate sterile neutrinos. Finally, in Section \ref{conclu} we draw our conclusions.

\section{Light Neutrino Exchange in $(\beta \beta)_{0\nu}$-decay and its connection to beta decay and cosmology \label{lightneu}}

Below we review  the most stringent  constraints on  $T^{0\nu}_{1/2}$  for the isotopes 
of interest $^{76}\rm{Ge}$, $^{136}\rm{Xe}$, 
$^{130}\rm{Te}$, 
$^{100}\rm{Mo}$ and $^{82}\rm{Se}$. 
All bounds are reported at 90$\%$ C.L. unless otherwise specified. 
\begin{enumerate}
\item
  The  claim of observation of $(\beta \beta)_{0\nu}$-decay  by H. V. Klapdor-Kleingrothaus and collaborators for the $^{76}\rm{Ge}$ isotope  
  corresponds to the half-life: $T^{0\nu}_{1/2}(^{76}\rm{Ge})=2.23^{+0.44}_{-0.31} \times 10^{25} \rm{yrs}$ (the range correspond to 68$\%$ C.L.) \cite{klapdor}. This has been challenged  by the previous results  from Heidelberg-Moscow \cite{hei}  and by the recent result   from GERDA \cite{gerda}. The   lower limit of  half-life of $^{76}\rm{Ge}$ that comes from GERDA \cite{gerda} is  $T^{0\nu}_{1/2}(^{76}\rm{Ge})>2.1\times 10^{25}\, \rm{yrs}$.  When combined with the limits from Heidelberg-Moscow (HDM) \cite{hei} 
and IGEX \cite{igex} experiments,  the  limit is $T^{0\nu}_{1/2}(^{76}\rm{Ge})>3.0\times 10^{25}\, \rm{yrs}$. Note that, as pointed out in Ref.~\cite{Klapdor-Kleingrothaus:2013cja, Dev:2013vxa}, the individual as well as the combined limit from GERDA does  not conclusively rule out the positive claim \cite{klapdor}.
\item
 The  bounds from EXO-200 \cite{Auger:2012ar} and KamLAND-Zen \cite{Gando:2012zm}
 experiments for  $^{136}\rm{Xe}$ are 
$T^{0\nu}_{1/2}(^{136}\rm{Xe})>1.6\times 10^{25}\, \rm{yrs}$ and 
$T^{0\nu}_{1/2}(^{136}\rm{Xe})>1.9\times 10^{25}\, \rm{yrs}$,  respectively. Combining the  two, the  lower limit becomes     
$T^{0\nu}_{1/2}(^{136}\rm{Xe})>3.4\times 10^{25}\, \rm{yrs}$ \cite{Gando:2012zm}.
\item
The bound   on the half-life of $^{130}\rm{Te}$ coming from CUORICINO  is  $T^{0\nu}_{1/2}(^{130}\rm{Te})>2.8\times 10^{24}\, \rm{yrs}$
\cite{cuo}.
\item
The lower limit on half-life of $^{100}\rm{Mo}$ from NEMO 3 is  $T^{0\nu}_{1/2}(^{100}\rm{Mo})>1.1\times 10^{24}\, \rm{yrs}$  \cite{nemo3}.
\item
The half-life of $^{82}\rm{Se}$ is bounded from below as  $T^{0\nu}_{1/2}(^{82}\rm{Se})>3.6\times 10^{23}\, \rm{yrs}$ \cite{nemo3}. 
\end{enumerate}
Among these different bounds, those on the half-life 
for $^{76}\rm{Ge}$ and $^{136}\rm{Xe}$ are in particular quite stringent. As pointed out in Ref.~\cite{Dev:2013vxa}, the claim of  
observation of $(\beta \beta)_{0\nu}$-decay in $^{76}\rm{Ge}$ is 
compatible with the individual limits from KamLAND-Zen and EXO-200 for few NME calculations, and it is in contradiction with the combined bound  for most of the NME calculations, except 
of the calculation corresponding to Ref.~\cite{engel}. 
For the discussion on the  mutual compatibility between the positive claim \cite{klapdor} and 
the bounds on the half-lives,  see also Ref.~\cite{faess}. It should be noted that, for a given value of $m^{\nu}_\mathrm{ee}$, the predicted value of the half-life $T^{0\nu}_{1/2}$ depends strongly on the NME uncertainty. Taking this variation into account, 
the correlation between half-lives for two different isotopes can be used to test the positive claim \cite{klapdor},  as it has been done in Refs.~\cite{Gando:2012zm,Dev:2013vxa}.

If  light neutrinos are  Majorana  particles \cite{Majorana}, they will  mediate  neutrinoless double beta decay  \cite{0nu2beta-old}. The observable in $(\beta \beta )_{0\nu}$-decay is the ee element of the mass matrix $|m^{\nu}_{\mathrm{e}\mathrm{e}}|$, 
known as the effective Majorana mass parameter of neutrinoless double beta decay, see e.g. Ref.~\cite{visreview, vuso, vis-fer, silvia0nu2beta, steve}. Explicitly
written in terms of the elements of the PMNS mixing matrix, this reads
\begin{equation}
m^\nu_{\mathrm{ee}} = m_1c_{12}^2c_{13}^2+m_2s_{12}^2c_{13}^2e^{2i\alpha_2}+m_3s_{13}^2e^{2i(\alpha_3+\delta)} \, , 
\label{mnuee}
\end{equation}
where $\alpha_{2,3}$ are the Majorana phases and $\delta$ is the Dirac phase. The half-life $T^{0\nu}_{1/2}$ of 
$(\beta \beta)_{0\nu}$-decay
 and the effective mass $m^{\nu}_\mathrm{ee}$ are related through the nuclear matrix element $\mathcal{M}_{\nu}$, the phase-space factor
$G_{0\nu}$ and electron mass $m_\mathrm{e}$ as
\begin{eqnarray}
\frac{1}{T_{1/2}^{0\nu}} 
= G_{0\nu}|{\cal M}_\nu|^2\left|\frac{m^\nu_\mathrm{ee}}{m_\mathrm{e}}\right|^2\,. 
\label{half1}
\end{eqnarray}
In Fig.~\ref{figsm} we show  the variation of $|m^{\nu}_\mathrm{ee}|$ with the lightest neutrino mass  $m_{\rm{lightest}}$, where we have used the $3 \sigma$ range of oscillation parameters from \cite{oscparam}.
 The blue and green areas correspond to $\alpha_{2,3}$ taking CP conserving values, while the red regions correspond to 
the violation of the CP symmetry.  
The dashed and dotted horizontal purple lines represent the required effective mass that will saturate the GERDA and GERDA+HDM+IGEX limits,  respectively \cite{gerda}. The orange lines correspond to the positive claim  (90$\%$ C.L.)\cite{klapdor}. The bands  represent the  NME uncertainty, taken from the compilation in Ref.~\cite{Dev:2013vxa}.
As the plot suggests, a measurement of $|m^{\nu}_\mathrm{ee}|$ will  give 
information on masses correlated with the CP violating phases, under the assumption
that light neutrino exchange is the only underlying mechanism in 
$(\beta \beta)_{0\nu}$-decay. 
%%%%%
\begin{figure}
\centering
\includegraphics[width=0.70\textwidth, angle=0]{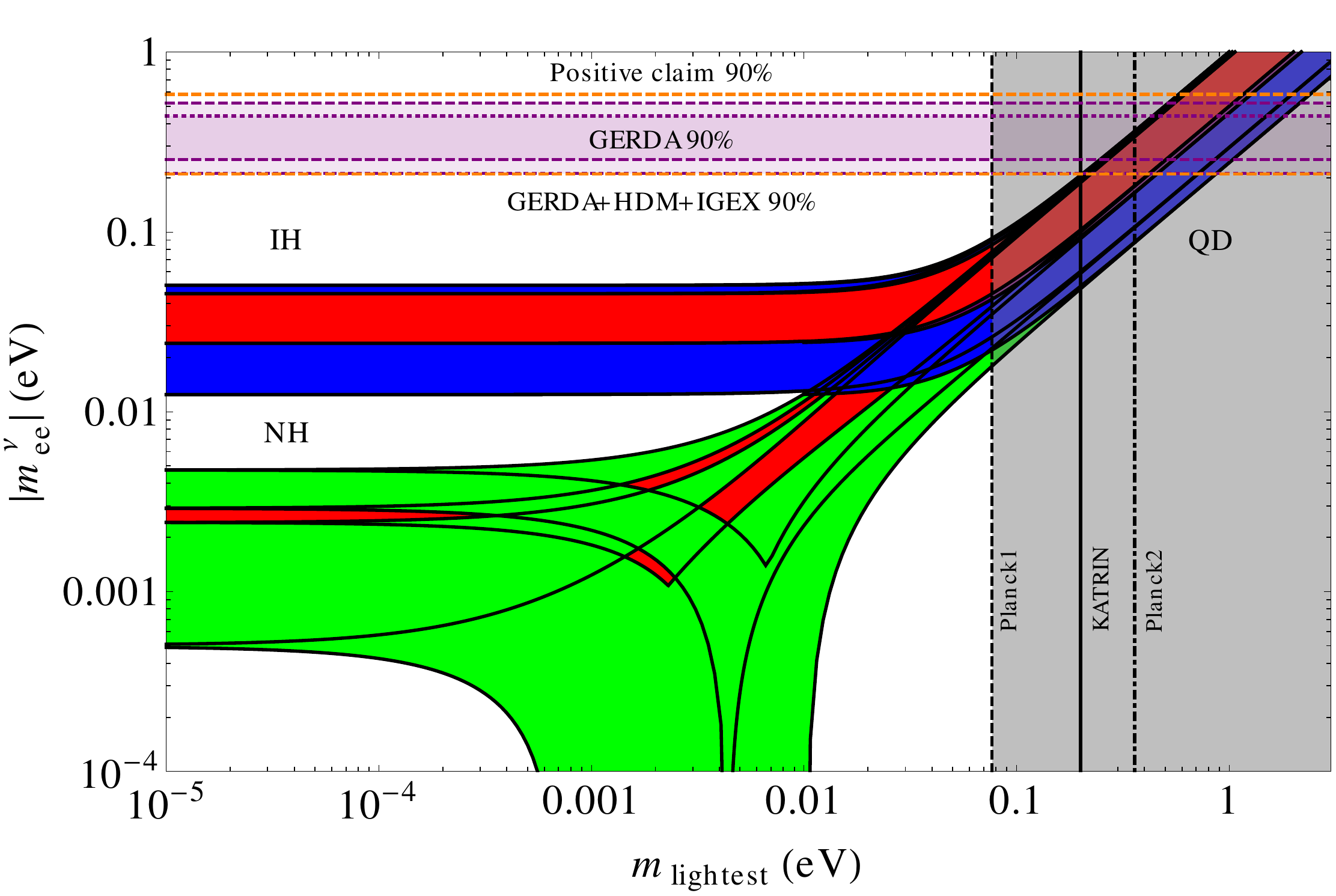}
\caption{{Variation of the effective mass $|m^{\nu}_\mathrm{ee}|$ with the lightest neutrino mass $m_{\rm{lightest}}$.  The horizontal purple lines represent the required  $|m^{\nu}_\mathrm{ee}|$ that will saturate the  limits of half-lives of $^{76}\rm{Ge}$ from GERDA \cite{gerda}. The purple band corresponds to the  NME uncertainty taken from the compilation in Ref.~\cite{Dev:2013vxa}. The orange lines correspond to the ranges of $|m^{\nu}_\mathrm{ee}|$ for which the half-life of $^{76}\rm{Ge}$ is  in 
agreement with the positive claim  ($90\% $ C.L.) \cite{klapdor}. The vertical black solid  line represents the KATRIN  sensitivity \cite{katrin}. The  dashed and dot-dashed    vertical lines represent the limits obtained from cosmology  $m_{\Sigma}=0.23 $ eV and $m_{\Sigma}=1.08$  eV \cite{Ade:2013lta}. }}
\label{figsm}
\end{figure}

In addition, the  light neutrino mass is also bounded  
from beta decay studies 
as well as from cosmology. The mass probed in beta-decay is 
$m_{\beta}=\sqrt{\Sigma_i |U^2_{ei}| m^2_i}$ { \cite{bp}} and the present  $95\%$~C.L. limit  on this observable is $m_{\beta}< 2.3~ \rm{eV}$ from MAINZ
\cite{mainz} and $m_{\beta}< 2.1 \, \rm{eV}$ from Troitsk \cite{troitsk}  collaborations,  respectively. This bound 
can be improved by 
one order of magnitude down to $m_{\beta}<0.2$ eV  from the beta decay experiment KATRIN \cite{katrin}, which is currently under commissioning. The sum of light neutrino masses $m_{\Sigma}=\Sigma_i m_i$ is constrained from cosmology. In the quasi-degenerate  
regime $m_{\rm{lightest}}> \sqrt{\Delta m^2_{\rm{atm}}}$, 
that is of particular interest for  $(\beta \beta)_{0\nu}$-decay, beta decay as well  cosmological searches, 
we have ${\Sigma_i m_i}/{3} \sim m_{\rm{lightest}} \sim  m_{\beta} \geq m^{\nu}_\mathrm{ee}$. The recent upper 
bounds on the sum of light neutrino masses coming from  Planck \cite{Ade:2013lta},
 which we consider in our studies, are the following: i) $ m_{\Sigma}<0.23$ eV, derived 
from the Planck+WP+highL+BAO data (Planck1) at $95\%$ C.L. and ii)   $ m_{\Sigma}<1.08$ eV from  Planck+WP+highL (AL) (Planck2) 
at $95 \%$ C.L. \cite{Ade:2013lta}. {As pointed out  in 
Refs.~\cite{fogli, MSV, msvmoriond, Dev:2013vxa}  and is evident from Fig.~\ref{figsm}}, after imposing the 
bounds from cosmology (assuming standard cosmology), the light neutrino contribution 
itself can not satisfy the claim in \cite{klapdor} or saturate the current
bounds \cite{hei, Gando:2012zm, gerda}.

%%%%%%%%%%%%%%%%%%%%%%%%%%%%%%%%%%%%%%%%%%%%%%%

\section{Sterile neutrino exchange in $(\beta \beta )_{0\nu}$-decay \label{largesterile}}

Sterile neutrinos 
can also give large contributions to neutrinoless double beta decay as analyzed in detail  in Refs.~\cite{f2005, ibarra, MSV, msvmoriond, Pascoli,  blennow, meroni2013}.
We assume here  sterile neutrinos~\footnote{For simplicity, we call the massive states mainly in the sterile neutrino direction simply "sterile neutrinos" as commonly done in the literature.} with a mass $M_i$ and which mix with $\nu_e$. The half-life $T^{0\nu}_{1/2}$ is \cite{kov, MSV}
\bea
\frac{1}{T^{0\nu}_{1/2}}=K_{0\nu} \left| \Theta^2_{\mathrm{e}i}\frac{M_i}{p^2-M^2_i}\right|^2,
\label{st}
\eea
where $K_{0\nu} \equiv G_{0\nu}(m_\mathrm{p} \mathcal{M}_N)^2$ and  
$p^2=-m_{\mathrm{e}}m_{\mathrm{p}}\frac{\mathcal{M}_N}{\mathcal{M}_{\nu}}$. Here  $\mathcal{M}_{\nu}$ is 
the NME for the light neutrino exchange and $\mathcal{M}_N$ is   for the 
heavy neutrino exchange, $|p|\sim 100 $ MeV is the exchanged momentum scale 
in $(\beta \beta)_{0\nu}$-decay,  $\Theta_{\rm{e}i}$ is the active-sterile neutrino mixing and $m_\mathrm{p}$ is the mass of the proton. In the subsequent discussions,  we denote $\Theta_{\rm{e}i}$ by $U_{\rm{e}{4_i}}$ and $M_i$ by $m_{4_i}$ for light sterile neutrinos, i.e.
when $M^2_i<< |p^2|$. For the heavy sterile case {$M^2_i>> |p^2|$}, and we denote them by $ V_{\mathrm{e}N_i}$ and $ M_{N_i} $, respectively. 
 For  light sterile neutrinos
 the above equation simplifies to
\bea
\frac{1}{T^{0\nu}_{1/2}} \simeq G_{0\nu} \mathcal{M}^2_{\nu}\left|\frac{ U^2_{\mathrm{e} 4_i} m_{4_i}} {m_\mathrm{e}}\right|^2,
\label{stl}
\eea
while for the heavy sterile one we have
\bea
\frac{1}{T^{0\nu}_{1/2}} \simeq G_{0\nu} \mathcal{M}^2_N \left| \frac{V^2_{\mathrm{e} {N_i}}m_\mathrm{p}}{M_{N_i}}\right|^2.
\label{sth}
\eea
Using the above equations and  the recent result   from GERDA \cite{gerda}, 
we derive the bound on the active-sterile mixing angle, assuming only one light or heavy sterile neutrino participates 
in $(\beta \beta)_{0\nu}$-decay. In all our subsequent analysis, we use the  values of NMEs $\mathcal{M}_{\nu}$ and 
$\mathcal{M}_N$ from Ref.~\cite{petcov}, corresponding to the axial vector cut-off $g_A=1.25$. We use the phase-space for 
$^{76}\rm{Ge}$: $G^{\rm{Ge}}_{0\nu}=5.77 \times 10^{-15} \, \rm{yrs}$ \cite{Kotila:2012zza}. In Fig.~\ref{figl}, we show 
the upper bound on the active-light sterile neutrino mixing angle $|U_{\mathrm{e}4}|^2$ from $(\beta \beta)_{0\nu}$-decay, that 
saturates the individual limit $T^{0\nu}_{1/2}= 2.1 \times 10^{25}$ yrs from GERDA \cite{gerda}. The gray region is due to the uncertainty introduced by the NME  $\mathcal{M}_{\nu}$  corresponding to the light  neutrino exchange. 
For comparison, we also show the other different bounds, first compiled in  Ref.~\cite{Atre:2009rg}. {For the mass of sterile neutrino $m_4< 1\,  \rm{MeV}$, the kink searches in $\beta$-decay spectrum is a sensitive probe of sterile neutrinos. The excluded regions with contours that are labelled by $^{187}\rm{Re}$, $^{3}\rm{H}$, $^{63}\rm{Ni}$, $^{35}\rm{S}$, $^{20}\rm{F}$ and $\rm{Fermi}_2$ refer to the bounds from kink searches \cite{Galeazzi:2001py, Hiddemann:1995ce, Holzschuh:1999vy, Holzschuh:2000nj, Deutsch:1990ut}.  
Note that,  in addition,  we have also included the bound coming from beta decay experiment of $^{64}\rm{Cu}$ 
\cite{Schreckenbach:1983cg}, which was not reported in Ref.~\cite{Atre:2009rg}. The reactor and solar  experiments  Bugey and Borexino \cite{Back:2003ae, Hagner:1995bn} 
are sensitive in the region $m_4 \sim $ few MeV. Exclusion contours have been drawn by looking into 
the decay of sterile neutrino into electron-positron pairs. On the other hand, 
for mass $m_4 >$ few MeV, the sensitive probe is the  peak search in $\pi \to e\nu$ \cite{Britton:1992pg}, where  the region inside the dot-dashed black contour is excluded.}  As can be seen from the figure, the bound on the active-light sterile neutrino mixing 
coming from $(\beta \beta)_{0\nu}$-decay  is the most stringent for most of the parameter spaces in  $U_{ \mathrm{e}4}-m_4$ plane. 
For the  mass  of the light sterile neutrino  $m_4 \lesssim 10^{-4}$ GeV,  the bounds from different beta decay searches 
are  close  to the ones from $(\beta \beta)_{0\nu}$-decay and possibly can be improved by the future beta decay experiments.
In the range  $10^{-4} \lesssim m_4 \lesssim 0.01$ GeV,  the bound from $(\beta \beta)_{0\nu}$-decay is the most stringent, 
while  around $m_4 \sim 0.1 $ GeV, the bound from peak searches, $\pi \to e\nu$  \cite{Britton:1992pg}, can almost compete with the bound from $(\beta \beta)_{0\nu}$-decay.

Similarly,  the upper limit on the mixing angle 
$|V_{\mathrm{e} N}|^2$ is shown in Fig.~\ref{figh}. The gray region is due to the uncertainty in the NME $\mathcal{M}_N$ corresponding to the heavy neutrino exchange. In addition, we also show  the other different bounds, from  Ref.~\cite{Atre:2009rg}.  {The regions inside the brown dot-dashed line is excluded from the beam dump experiment PS191 \cite{Bernardi:1987ek}. For mass of sterile neutrino $M_N \sim \mathcal{O}(100)$ MeV, the stringent bound is obtained from the  electron  spectrum in meson decay  $K \to e \nu$ decay \cite{knupeak}. For heavier masses $M_N \sim \mathcal{O}$(GeV), the 
$Z^0$ decays into sterile neutrinos can be used to obtain exclusion contours, labelled as DELPHI and L3 
\cite{Abreu:1996pa, Adriani:1992pq}.  See Ref.~\cite{Atre:2009rg} and the references therein for the detail description
 of other different bounds \cite{Badier:1986xz,Bergsma:1985is}.} Also in this case, for most of the parameter space, the $(\beta \beta)_{0\nu}$-decay gives the most stringent limit.  For the mass of the heavy sterile neutrino $M_N  \sim \mathcal{O}(100) $ MeV, the bound 
from  the beam dump experiment PS191  is competitive with the one from $(\beta \beta)_{0\nu}$-decay. 
 For the positive claim \cite{klapdor}, the results are very similar and we do not show the corresponding region  explicitly.

\begin{figure}[hbt]
\centering
\includegraphics[width=0.80\textwidth, angle=0]{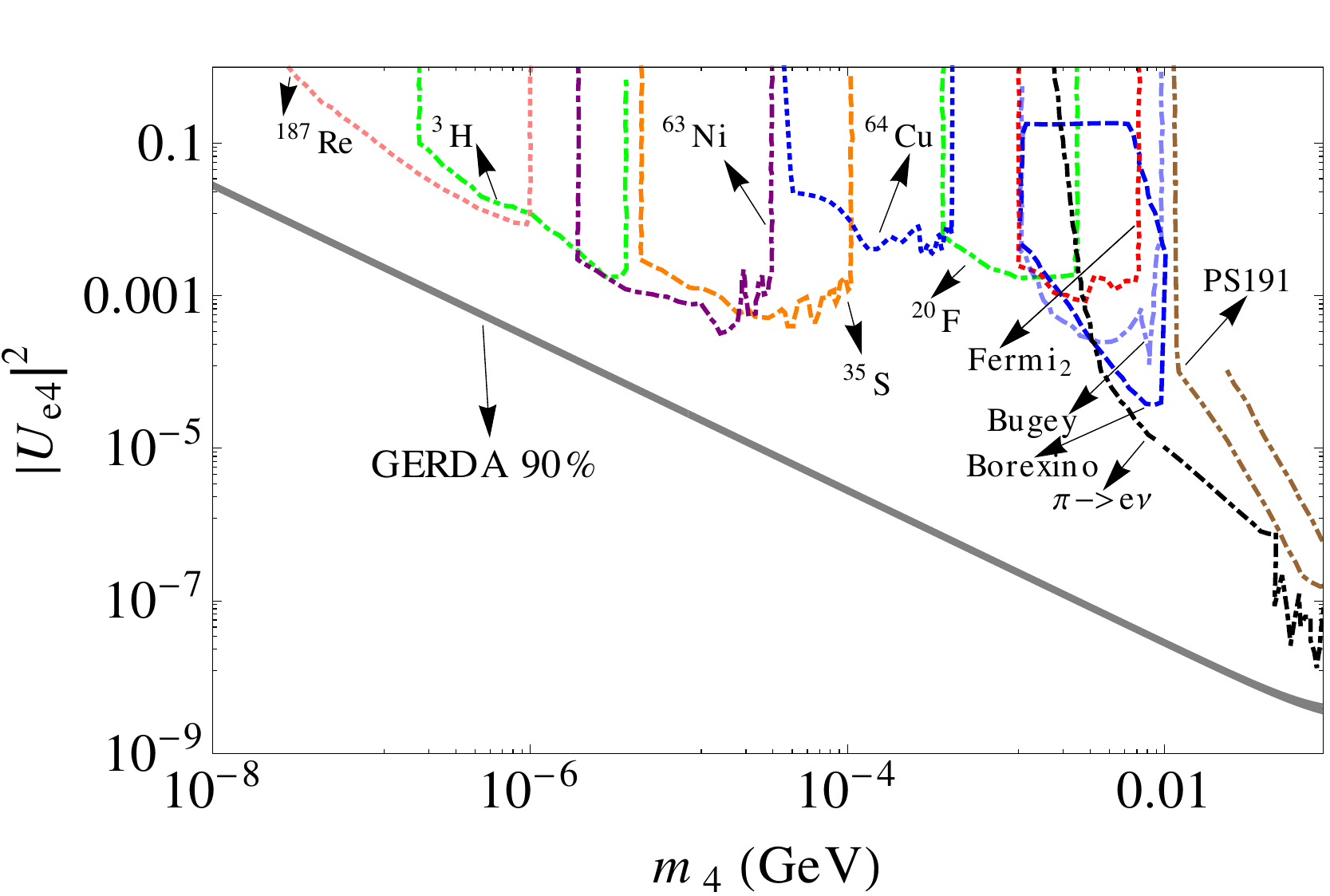}
\caption{Upper bound  of  $|U^2_{\mathrm{e}4}|$ as a function of $m_4$ from the limit on the half-life from GERDA \cite{gerda}. 
The gray band is due to  the uncertainty on the NMEs.
For comparison, we  also show  the different bounds from  beta decay, solar and reactor experiments, peak search 
  and beam dump experiment, first compiled in Ref.~\cite{Atre:2009rg} . }
\label{figl}
\end{figure}
\begin{figure}[hbt]
\includegraphics[width=0.80\textwidth, angle=0]{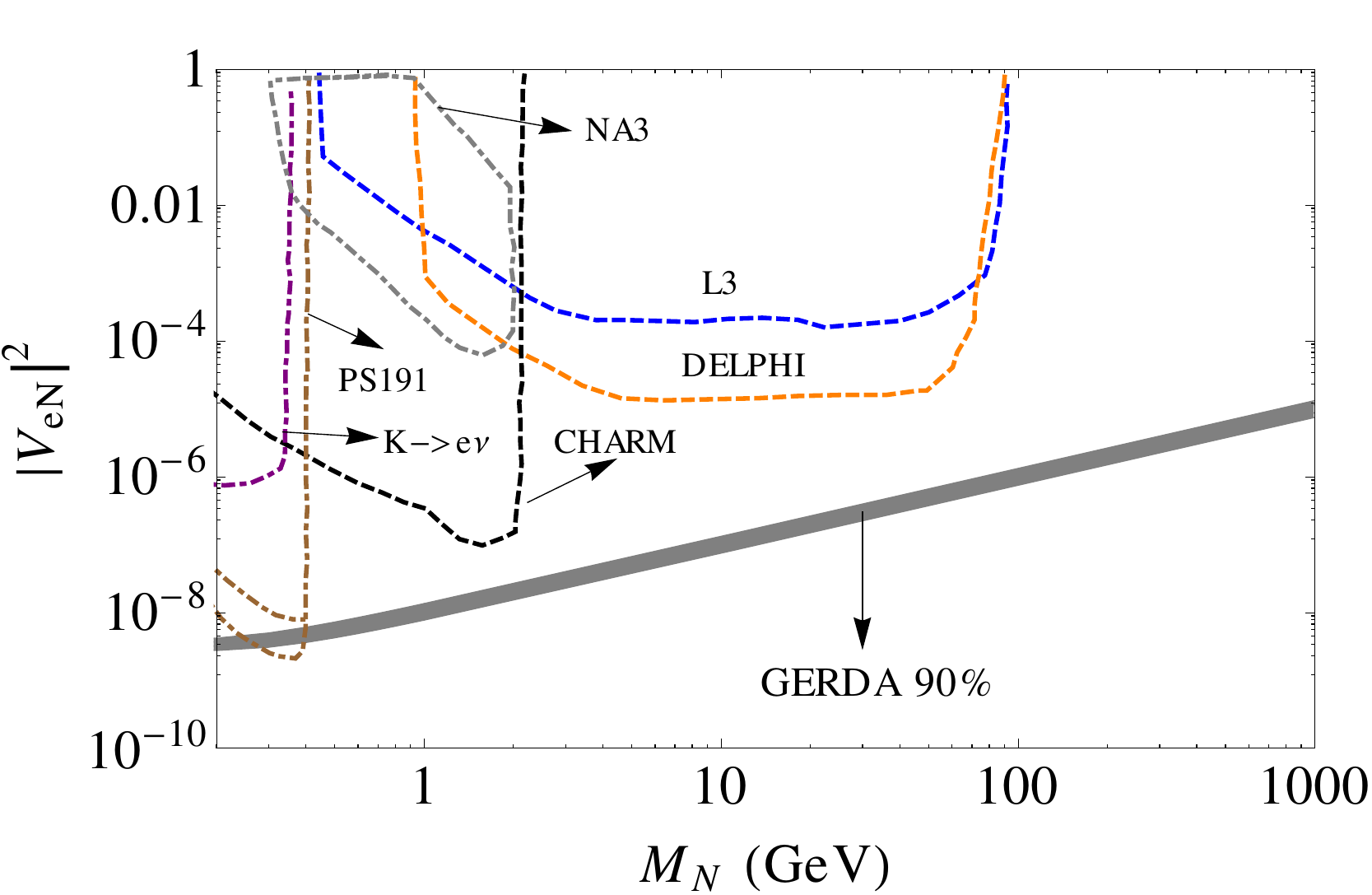}
\caption{The same as Fig.~\ref{figl} but for $|V^2_{\mathrm{e} N}|$ versus $M_N$. See text for the detail.}
\label{figh}
\end{figure}

%%%%%%%%%%%%%%%%%%%%%%%%%%%%%%%%%%%%%%%%%%%%%%%%%%%%%%%%%%%%

\section{Cancellations among different contributions in $(\beta \beta)_{0\nu}$-decay \label{cance}}

 The   discussion of the previous section on the effective Majorana mass relies on the  assumption that either the light or heavy neutrino exchange is the only underlying mechanism in $(\beta \beta)_{0\nu}$-decay. However, in an extension of the standard model leading to light Majorana masses, the lepton number violating mechanism responsible for it will also contribute to neutrinoless double beta decay directly and could potentially interfere with the light neutrino one. Below we consider this possibility in detail.
This is of particular interest, as it can solve the mutual inconsistency 
between  the positive claim \cite{klapdor} and the results from KamLAND-Zen \cite{Gando:2012zm}.
%%%%

If more than one mechanism is operative at $(\beta \beta)_{0\nu}$-decay, the 
half-life $T^{0\nu} _{1/2}$  of $(\beta \beta)_{0\nu}$-decay  for a particular isotope will receive different contributions as
\be
\frac{1}{T^{0\nu} _{1/2}}=G_{0\nu}(|\eta^2_1|\mathcal{M}^2_1+|\eta^2_2| \mathcal{M}^2_2+
2 \cos \alpha\,  |\eta_1| |\eta_2| \mathcal{M}_1\mathcal{M}_2), 
\label{half-two}
\ee
where $G_{0\nu}$ is the phase-space factor, $\mathcal{M}_{1,2}$ are the NMEs for the two different  exchange processes. 
Here,  $\eta_{1}$ and $\eta_2$ are the two dimensionless quantities which 
contain all the information from  the particle physics parameters associated
with the two exchange mechanisms and  $\alpha$ is the relative phase factor between them. 
The different exchange mechanisms can be for  e.g. light neutrino and sterile neutrino exchange, 
or light neutrino and squark/gluino exchange. 
If a complete cancellation takes place between two exchange mechanisms, then the phase 
 $\cos\,\alpha=-1$ and $|\eta_1|\mathcal{M}_1=|\eta_2|\mathcal{M}_2$.  Consequently the half-life $T^{0\nu}_{1/2}$ 
 in Eq.~\ref{half-two} would be infinite, and this process in a specific nucleous 
would never be observed. However, this does not need to be the case for another isotope. Between two isotopes (A, B), if this cancellation is effective for isotope A, then the  half life for isotope B is 
\be
\frac{1}{T^{0\nu}_{1/2}({B})}=G^{{B}}_{0 \nu}|\eta^2_1|({\mathcal{M}_{1,B}}-  \frac{{\mathcal{M}_{1,A}}}{{\mathcal{M}_{2,A}}} {\mathcal{M}_{2,B}})^2, 
\label{ge}
\ee
where $\mathcal{M}_{1,A}$, $\mathcal{M}_{2,A}$ are the NMEs for the two exchange processes in isotope A and $\mathcal{M}_{1,B}$, $\mathcal{M}_{2,B}$ are for  isotope B.  As an example we consider the case when the cancellation is effective in $^{136}\rm{Xe}$. In this case,  the bound on half-life $T^{0\nu}_{1/2}>3.4 \times 10^{25} \, \rm{yrs}$ 
\cite{Gando:2012zm}
is automatically satisfied, irrespective of the absolute magnitude of $|\eta_{1,2}| $. 
Denoting the nuclear 
matrix elements for $^{76}\rm{Ge}$ and $^{136}\rm{Xe}$ by $\mathcal{M}_{1, \rm{Ge}}$, $\mathcal{M}_{2, \rm{Ge}}$ and  $\mathcal{M}_{1,\rm{Xe}}$, $\mathcal{M}_{2,\rm{Xe}}$   and the 
phase space of $^{76}\rm{Ge}$ by $G^{\rm{Ge}}_{0\nu}$, the half-life of $^{76}\rm{Ge}$ is
\be
\frac{1}{T^{0\nu}_{1/2}(^{76}\rm{Ge})}=G^{\rm{Ge}}_{0 \nu}|\eta^2_1|(\mathcal{M}_{1,\rm{Ge}}-  \frac{\mathcal{M}_{1,\rm{Xe}}}{\mathcal{M}_{2,\rm{Xe}}} \mathcal{M}_{2,\rm{Ge}})^2.
\label{ge}
\ee
The value  of $|\eta_1|$ that saturates the lower limit of half-life from GERDA \cite{gerda} and  GERDA+HDM+IGEX \cite{gerda} are 
\begin{equation}
|\eta_1| \leq \frac{(2.87,2.40) \times 10^{-6}}{{\big|({\mathcal{M}_{1, \rm{Ge}}}-  \frac{{\mathcal{M}_{1,\rm{Xe}}}}{{\mathcal{M}_{2,\rm{Xe}}}} {\mathcal{M}_{2,\rm{Ge}}})\big|}},
\label{eta1}
\end{equation}
 while the range of $|\eta_1|$ that satisfies the positive claim ($90\%$ C.L.)  in \cite{klapdor}   is 
\begin{equation}
|\eta_1|=\frac{(2.42-3.18) \times 10^{-6}}
{{\big|({\mathcal{M}_{1, \rm{Ge}}}-  \frac{{\mathcal{M}_{1,\rm{Xe}}}}{{\mathcal{M}_{2,\rm{Xe}}}} {\mathcal{M}_{2,\rm{Ge}}})\big|}}.
\label{eta2}
\end{equation}
 As stressed before, note that, the individual or the combined limit from GERDA \cite{gerda} does not conclusively rule  out the positive claim in \cite{klapdor}. Hence, in addition to the GERDA,  GERDA+HDM+IGEX limits \cite{gerda}, we also carry out the discussion on the  positive claim \cite{klapdor}. If the above mentioned cancellation is operative for $^{136}\rm{Xe}$,  it would be possible 
to automatically satisfy  the bounds obtained by EXO-200, KamLAND-Zen collaboration \cite{Auger:2012ar, Gando:2012zm} 
for $^{136}\rm{Xe}$ isotope and yet 
to satisfy the  claim in \cite{klapdor}, irrespective of any NME uncertainty. Hence, it is possible to reconcile any mutual conflict  between 
the results of $^{136}\rm{Xe}$ and $^{76}\rm{Ge}$.

\subsection{Light active and heavy sterile neutrinos \label{heavy}}

We first discuss the case when the two interfering mechanisms correspond  to light active
and heavy sterile neutrino exchange. We also include the discussion when the cancellation is operative 
between light neutrino exchange and squark/gluino exchange mechanisms, 
for e.g. in R-parity violating supersymmetry.

\begin{figure}
\centering
\includegraphics[width=0.70\textwidth, angle=0]{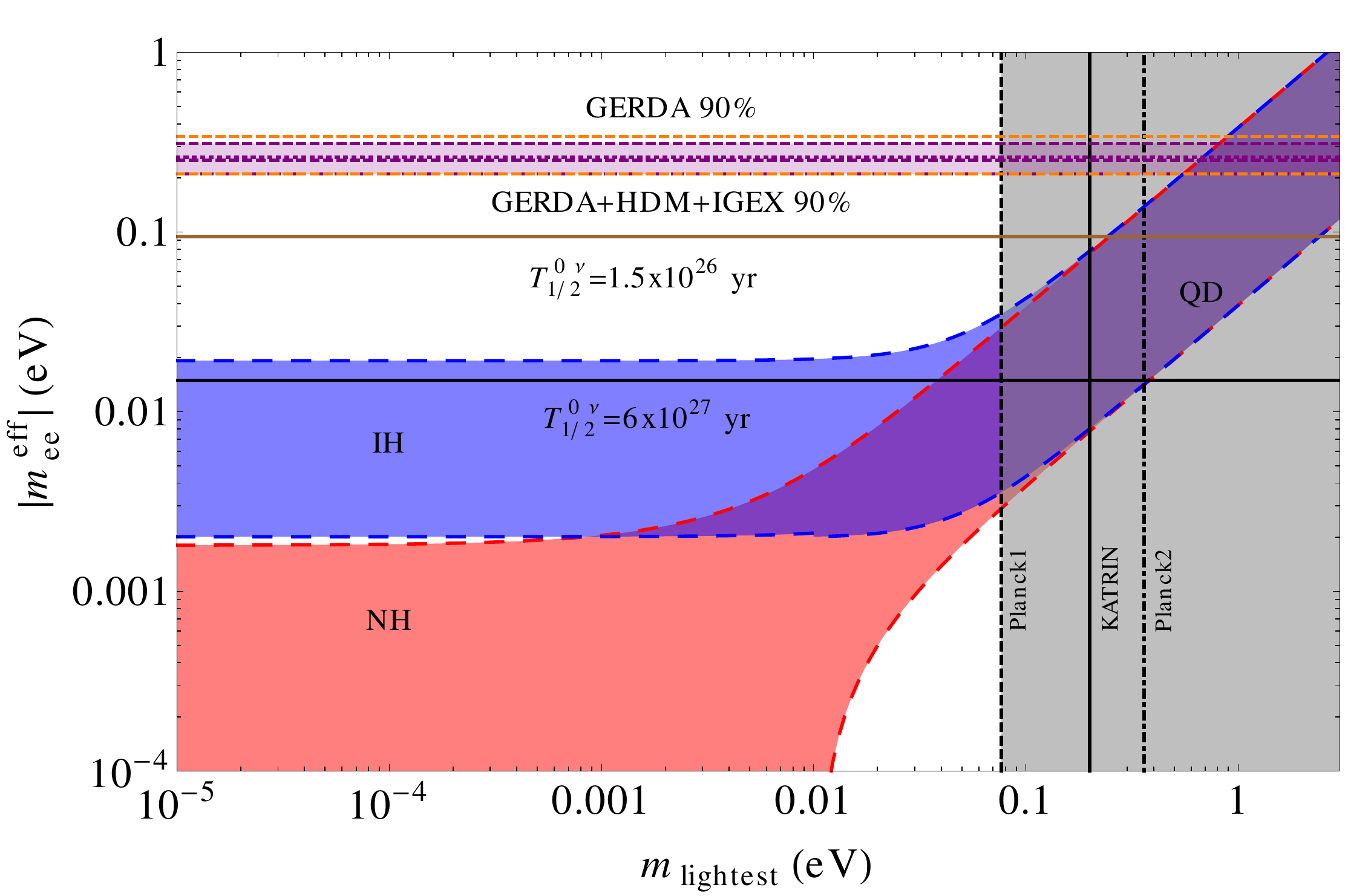}
\caption{Variation of redefined effective mass $|m^{\rm{eff}}_\mathrm{ee}|$ with the lightest neutrino mass $m_{\rm{lightest}}$ for $^{76}\rm{Ge}$. The effect of cancellation is operative between light active and heavy sterile neutrino. 
 The horizontal purple lines  represent the required  $|m^{\rm{eff}}_\mathrm{ee}|$ that will saturate  the limits  of half-lives  from GERDA \cite{gerda}. 
The horizontal dashed orange lines represent the ranges of $|m^{\rm{eff}}_\mathrm{ee}|$ for which 
the half-life for $^{76}\rm{Ge}$ is  in 
agreement with the positive claim ($90\% $ C.L.) \cite{klapdor}.
The vertical black solid line represents the future sensitivity of KATRIN \cite{katrin}. {The  dashed and dot-dashed    vertical lines represent the limits obtained from cosmology \cite{Ade:2013lta}.} The horizontal brown and black lines show the future sensitivity of the effective mass for $^{76}\rm{Ge}$, assuming a half-life $T^{0\nu}_{1/2}=1.5 \times 10^{26}\, \rm{yr}$ ( GERDA Phase-II) 
\cite{gerdafuture} and $T^{0\nu}_{1/2}=6 \times 10^{27}\, \rm{yr}$  \cite{werner-rev},  respectively. }
\label{fig1}
\end{figure}

 First,  we study the case of light active  neutrinos $\nu_i$ and heavy sterile $N_j$ with mass $M_{N_j}$,  larger than the typical 
momentum exchange  $|p|$ in $(\beta \beta)_{0\nu}$-decay: $M^2_{N_j}>> |p^2| \sim (100)^2 \, 
\rm{MeV}^2 $. We consider maximum destructive interference between the two 
mechanisms, i.e. $\cos\, \alpha=-1$. A cancellation in isotope A  will lead to the following relation,  
\begin{eqnarray}
|\eta_N|=|\eta_{\nu}| \frac{\mathcal{M}_{\nu,A}}{\mathcal{M}_{N,A}}.
\label{canc}
\end{eqnarray}

 Here,  we have replaced $\eta_{1,2}$ of the previous section  by $\eta_{\nu,N}$, 
respectively, where $\eta_{\nu}$ correspond to light neutrino exchange and 
$\eta_N$ correspond to the heavy sterile neutrino exchange. The  nuclear matrix elements $\mathcal{M}_{1,A}$ and $\mathcal{M}_{2,A}$ in this case correspond to light and heavy neutrino exchange and have been denoted as $\mathcal{M}_{\nu,A}$ and $\mathcal{M}_{N,A}$,   respectively.} In the above, the particle physics dimensionless parameters 
$\eta_{\nu}$ and $\eta_N$ are given by 
\begin{eqnarray}
\eta_{\nu}&=&\frac{m^{\nu}_\mathrm{ee}}{m_\mathrm{e}},\\ 
\eta_N &=& \sum_j \frac{V^2_{\mathrm{e}j}m_\mathrm{p}}{M_{N_j}}.
\end{eqnarray} 
The half life for any other isotope B is 
predicted to be 
\be
\frac{1}{T^{0\nu}_{1/2}({B})}=G^{{B}}_{0 \nu}\left|\frac{m^{\nu}_\mathrm{{ee}}}{m_\mathrm{e}}\right|^2\mathcal{M}^2_{\nu,B}
\left(1- \frac{\mathcal{M}_{\nu,A}}{\mathcal{M}_{N,A}} \frac{\mathcal{M}_{N,B}}{\mathcal{M}_{\nu,B}}\right)^2.
\label{lh}
\ee
It can be rewritten in terms of an effective mass, where 
 the {\it redefined effective mass} is 
\be
\left|m^{\rm{eff}}_{\mathrm{ee}}\right|=\left|m^{\nu}_\mathrm{ee}(1- \frac{\mathcal{M}_{\nu,A}}{\mathcal{M}_{N,A}} \frac{\mathcal{M}_{N,B}}{\mathcal{M}_{\nu,B}})\right|. %< \left|m^{\nu}_\mathrm{ee}\right|. 
\label{effm}
\ee
Hence,  if the light and heavy exchange contributions cancel each other for isotope A, for any other isotope B the effect would manifest itself by {increasing the half-life.} %lowering the effective mass. 
Below, as a relevant example, we again  focus  on  the case in which the cancellation is present   in $^{136}\rm{Xe}$ and we explore its effect on the half-life of  $^{76}\rm{Ge}$.

\begin{table}[h!]
\begin{center}
\begin{tabular}{c|c|c|c}\hline\hline
 NME & \multicolumn {3}{|c}{$|m^{\rm{eff}}_\mathrm{ee}|$ (eV)} \\ \cline{2-4}
 SRQRPA & GERDA & Combined &  Positive claim \\ \hline
Argonne intm & 0.31 &  0.26  & 0.26-0.34 \\
Argonne large & 0.27 & 0.23 & 0.23-0.30 \\
CD-Bonn & 0.29 & 0.24 & 0.24-0.32 \\
CD-Bonn & 0.25 & 0.21 & 0.21-0.28\\ \hline
\hline\hline
\end{tabular}
\end{center}
\caption{The upper limits of the redefined effective neutrino mass $|m^{\rm{eff}}_\mathrm{ee}|$ that saturate
the  lower limits  of  half-life of $^{76}\rm{Ge}$ from GERDA \cite{gerda} and GERDA+HDM+IGEX \cite{gerda}. The NMEs have been taken from  \cite{petcov}. Also shown are its required  ranges  corresponding to the positive claim 
($90\%$ C.L. ) \cite{klapdor}.
}
\label{tabsterile}
\end{table}

Using Eq.~\ref{lh}, the different values of redefined effective mass $|m^{\rm{eff}}_\mathrm{ee}|$ that 
is required to saturate the individual and combined limits of half-life   from GERDA \cite{gerda} and to satisfy the positive claim ($90\%$ C.L.)  \cite{klapdor} are given  in Table~\ref{tabsterile}. The {\it redefined effective mass} 
$|m^{\rm{eff}}_{ee}|$ is smaller than the true effective mass $|m^{\nu}_{ee}|$, as expected.  We  show the variation of the effective mass $|m^{\rm{eff}}_\mathrm{ee}| $ with 
the lightest neutrino mass scale $m_{\rm{lightest}}$ in Fig.~\ref{fig1}. The horizontal purple bands show the effect of NME uncertainties   and correspond to the two different ranges of required effective masses $ |m^{\rm{eff}}_\mathrm{ee}|=(0.25-0.31)$ eV (dashed purple band) and $|m^{\rm{eff}}_\mathrm{ee}|=(0.21-0.26)$ eV (dotted purple band) to saturate the  GERDA and GERDA+HDM+IGEX \cite{gerda}  limits,  respectively. 
The horizontal dashed  orange   lines represent the minimum and 
maximum of the required  ranges of  effective mass  $|m^{\rm{eff}}_\mathrm{ee}|=(0.21-0.34)\, \rm{eV}$ that satisfies the positive claim \cite{klapdor}. 
In both of the figures, the vertical  black solid line 
represents the future sentivity of KATRIN $m_{\rm{lightest}}<0.2$~eV \cite{katrin} and the other two vertical lines represent 
the bound $m_{\rm{lightest}}<0.077 \, \rm{eV}$ and $m_{\rm{lightest}}<0.36\, \rm{eV}$, following the two extreme bounds from 
Planck data set $m_{\Sigma}<0.23\, \rm{eV}$ (Planck1) and $m_{\Sigma}< 1.08 \, \rm{eV}$ (Planck2) \cite{Ade:2013lta},  respectively.

\begin{figure}
\centering
\includegraphics[width=0.490\textwidth, angle=0]{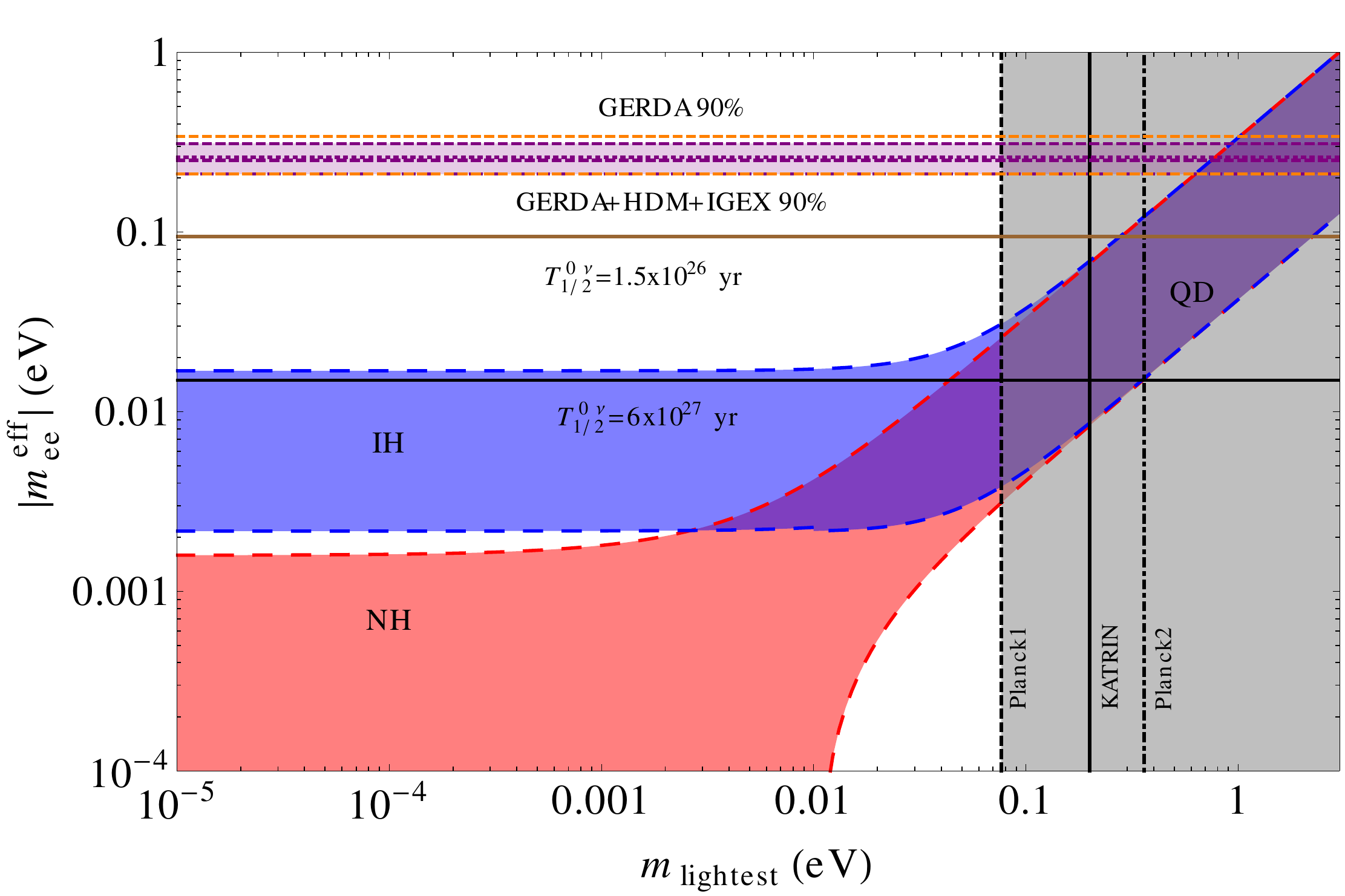}
\includegraphics[width=0.490\textwidth, angle=0]{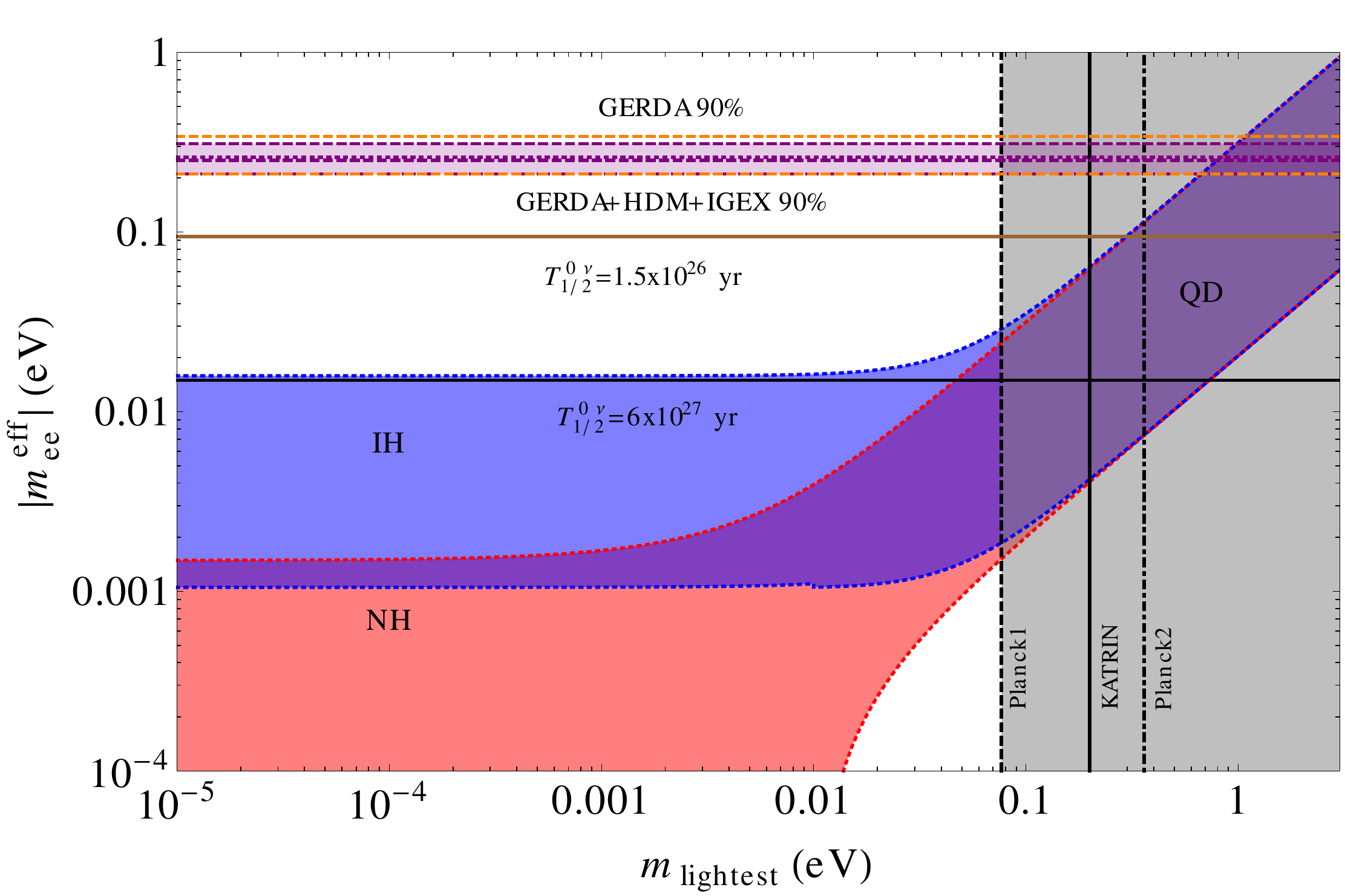}
\caption{Variation of the redefined effective mass $|m^{\rm{eff}}_\mathrm{ee}|$ with the lightest neutrino mass $m_{\rm{lightest}}$ for $^{76}\rm{Ge}$. Left Panel: the cancellation is effective between light neutrino and  gluino exchange.  The horizontal  purple lines represent the required $|m^{\rm{eff}}_\mathrm{ee}|$ that will saturate the limits from GERDA  \cite{gerda}. The  horizontal 
 dashed orange lines represent the ranges of $|m^{\rm{eff}}_\mathrm{ee}|$ for which the half-life for $^{76}\rm{Ge}$ is  in agreement positive claim   ($90\% $ C.L.) \cite{klapdor}. 
See text for details. Right panel: for the case when the cancellation is effective between light neutrino and squark exchange. All other descriptions remain same as in Fig.~\ref{fig1}. }
\label{fig2}
\end{figure}

 As can be seen from the figure, the effective mass $|m^{\rm{eff}}_\mathrm{ee}|$ 
can saturate the required values  only in the 
quasi-degenerate regime. However, this possibility can be   severely constrained by  the future    
sensitivity  from KATRIN \cite{katrin},
which does not depend on any particular physics model. In particular, for the bound $m_{\beta}<0.2 \, \rm{eV}$ from KATRIN \cite{katrin}, the effective mass  can not reach 
 the required value of $|m^{\rm{eff}}_\mathrm{ee}|$.
The bound from cosmology is  even more stringent compared to the case when 
the light neutrinos are the only mediators and therefore the tension between cosmology and the possible claim
in neutrinoless double beta decay is more severe.
We also show the future sensitivity for $^{76}\rm{Ge}$ by the horizontal brown and black lines that correspond to half-lives $T^{0\nu}_{1/2}=1.5 \times 10^{26} \, \rm{yrs}$ for GERDA Phase-II \cite{gerdafuture} and  $T^{0\nu}_{1/2}=6 \times 10^{27}\, \rm{yrs}$ \cite{werner-rev},  respectively.  {It is evident from Fig.~\ref{fig1}, that the effective mass can  saturate the future limit from GERDA Phase-II around $m_{\rm{lightest}} \sim 0.25$ eV. This possibility is unconstrained from the most stringent limit from Planck and marginally constrained by the future sensitivity of KATRIN. For the half-life 
$T^{0\nu}_{1/2}=6 \times 10^{27}\, \rm{yrs}$, the effective mass can  saturate the limit even for $m_{\rm{lightest}}$ as low as $10^{-5}$ \rm{eV}. This possibility is not at reach for future cosmological observations and beta decay experiments.

The cancellation between light contribution and heavy contribution can also be realized in other new physics scenarios, for  e.g. R-parity violating supersymmetry. In this framework, the gluino and squarks can give large contribution 
in $(\beta \beta)_{0\nu}$-decay. Below,  we discuss  the case  when the cancellation 
is effective between  light neutrino exchange and gluino/squark exchange. We denote the NMEs corresponding to the gluino exchange by $\mathcal{M}_{\lambda'}$ and the squark exchange by $\mathcal{M}_{\tilde{q}}$ and parameterize their contributions by $\eta_{{\lambda'}}$ and $\eta_{\tilde{q}}$, respectively. The detail description of $\eta_{{\lambda'}}$ and $\eta_{\tilde{q}}$ on the 
fundamental parameters of the theory has been described in detail in Ref.~\cite{f2010, allanach}, and we do not repeat them here. Like the previous case, 
the cancellation between light neutrino exchange and squark/gluino exchange in isotope A will result in a reduction 
of effective mass  for any other isotope. 
The left and right panels  of  Fig.~\ref{fig2} corresponds to the two different cases, when the cancellation is effective between light neutrino-gluino and light neutrino-squark exchanges,  respectively. The NMEs have been used from Ref.~\cite{petcov}. The horizontal dashed and dotted purple lines represent the required 
effective mass that will saturate GERDA and GERDA+HDM+IGEX \cite{gerda} limits. 
They have been derived using Eq.~\ref{lh}  and includes the  effect of cancellation in $^{136}\rm{Xe}$.  The  horizontal orange 
 lines correspond to  the required ranges of effective mass
$|m^{\rm{eff}}_\mathrm{ee}|$, that will satisfy the positive claim \cite{klapdor}. 

%%%%%%%%%%%%%%%%%%%%%%%%%%%

\subsection{Light and   heavy sterile neutrinos \label{light-heavy}}

The tension discussed above between cosmology and neutrinoless double beta decay can be avoided if, in addition to the heavy sterile neutrinos, 
we also have  light sterile neutrinos. 
The latter, depending 
on their mass and mixing, can give a large  contribution even compared to the light active ones and can 
infact saturate the required value of $|m^{\rm{eff}}_\mathrm{ee}|$. On the 
other hand, the bounds from cosmology 
is only relevant if the masses of the sterile neutrinos are very small $m_{4} \sim \rm{eV}$
and they were copiously produced in the Early Universe contributing to hot dark matter. For heavier masses, $m_4 > \mathrm{keV}$, the mixing angles of interest are very large and would lead to fast decays of sterile neutrinos and consequently to no bounds from cosmology. 
Hence, adding light sterile 
neutrinos in addition to heavy sterile ones  can solve the mutual inconsistency between 
the positive claim in \cite{klapdor} and KamLAND-Zen \cite{Gando:2012zm},  can saturate the upper limits of 
effective masses for  GERDA and GERDA+HDM+IGEX  \cite{gerda} and  can be in accordance with  
the bounds coming from  cosmology.  
Here, we study in detail this case .

We assume both Majorana light sterile neutrinos $\nu_{4_k}$ with mass $m^2_{4_k}<<|p^2| \sim (100)^2\, \rm{MeV}^2$   
and heavy sterile neutrinos $N_j$ with mass $M^2_{N_j}>> |p^2| \sim (100)^2\,  \rm{MeV}^2$. 
In this case, the half-life  of  any isotope is
\be
\frac{1}{T^{0\nu}_{1/2}}=G_{0\nu}(|\eta^2_{l}|\mathcal{M}^2_{\nu}+|\eta^2_N| \mathcal{M}^2_N+2 \cos\,{\alpha}\, |\eta_l| \,|\eta_N| \, \mathcal{M}_l\, \mathcal{M}_N), 
\ee
where the parameters $\eta_l$ and $\eta_N$ correspond to the contributions from light and heavy neutrinos as
\begin{equation}
\eta_l=\frac{(\Sigma_i m_i U^2_{\mathrm{e}i}+ \Sigma_{k} m_{4_k} U^2_{\mathrm{e}4_k})}{m_\mathrm{e}}, \, \, \eta_N=\sum_j \frac{m_\mathrm{p} V^2_{{\mathrm{e}N}_j}}{M_{N_j}}.
\end{equation}

\begin{figure}[hbt]
\centering
\includegraphics[width=0.80\textwidth, angle=0]{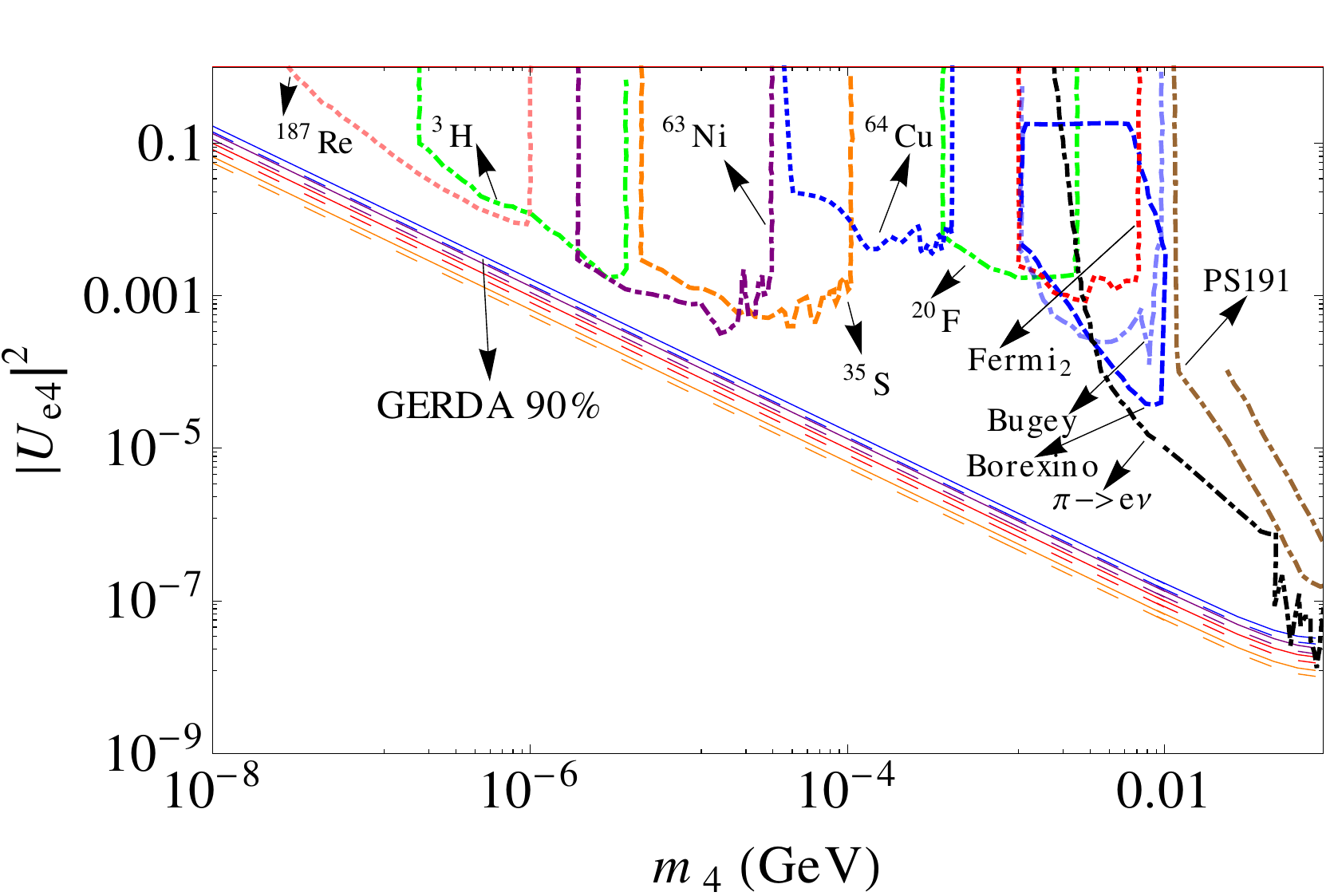}
\caption{Upper bounds  of  $|U^2_{e4}|$ that saturate the limits from  GERDA \cite{gerda}. 
The different color coding corresponds to the NME uncertainty. See text for details. 
 For comparison, we  also show  the different bounds from  beta decay, solar and reactor experiments, peak search and beam dump experiment, first compiled in Ref.~\cite{Atre:2009rg}.} 
\label{fig3}
\end{figure}

For  simplicity we consider the case in which only one light sterile and one heavy sterile neutrinos
are present. If the cancellation between light and heavy neutrino contribution is effective for isotope A, then  following the discussions of previous sections,
$\eta_l$ and $\eta_N$ are related as $|\eta_N|=|\eta_l|\frac{\mathcal{M}_{\nu,A}}{\mathcal{M}_{N,A}}$. For any other isotope B, 
the {\it redefined
effective mass} $m^{\rm{eff}}_\mathrm{ee}$ is %, given by 
\be
m^{\rm{eff}}_\mathrm{ee}=(m^{\nu}_\mathrm{ee}+  m_{4} U^2_{\mathrm{e}4})\times (1- \frac{\mathcal{M}_{\nu,A}}{\mathcal{M}_{N,A}} \frac{\mathcal{M}_{N,B}}{\mathcal{M}_{\nu,B}}).
\label{meffsterile2}
\ee
In the above we have dropped the generation index and $m_4$ denotes the mass of the light sterile neutrino, while $U_{\mathrm{e}4}$ is the 
active-light sterile mixing. We again assume a  cancellation for $^{136}\rm{Xe}$ and we examine its  
implications on  $^{76}\rm{Ge}$. From Table~\ref{tabsterile},  it is evident
 that to satisfy/saturate either  the positive claim \cite{klapdor} or the limits from GERDA \cite{gerda}, a large effective 
mass  $|m^{\rm{eff}}_\mathrm{ee}| \sim \mathcal{O}(0.1)$ eV is required. We denote  the  limiting values of  effective masses 
$|m^{\rm{eff}}_\mathrm{ee}|$ of  Table~\ref{tabsterile} by  $\kappa$ for GERDA, GERDA+HDM+IGEX \cite{gerda}  and  
the minimum and maximum values of the required  $|m^{\rm{eff}}_\mathrm{ee}|$ 
by $\kappa_1$ and $\kappa_2$ for the positive claim \cite{klapdor}. 
Following the stringent constraint from cosmology,  %and future beta decay experiment, 
 the effective neutrino mass $|m^{\nu}_\mathrm{ee}|$ corresponding to the  light neutrino exchange 
is extremely small $|m^{\nu}_\mathrm{ee}|<  0.09 $ eV  (see Fig.~\ref{figsm}) 
and we will neglect it in the following. Hence,  if the total
contribution  saturates the limits from GERDA and  
GERDA+HDM+IGEX  \cite{gerda}, 
  the active-light sterile neutrino mixing  
$|U^{2}_{\mathrm{e}4}|$ is bounded as follows
\begin{equation}
 |U^2_{\mathrm{e}4}|\lesssim \frac{\kappa}{m_4}\frac{1}{\left|(1- \frac{\mathcal{M}_{\nu,\rm{Xe}}}{\mathcal{M}_{N,\rm{Xe}}} \frac{\mathcal{M}_{N,\rm{Ge}}}{\mathcal{M}_{\nu,\rm{Ge}}})\right|} ~.
\label{boundUgerda}
\end{equation}
On the other hand, in order to explain the positive claim in 
Ref.~ \cite{klapdor}   we need,
\begin{equation}
\frac{\kappa_1}{m_4}\frac{1}{\big|(1- \frac{\mathcal{M}_{\nu,\rm{Xe}}}{\mathcal{M}_{N,\rm{Xe}}} \frac{\mathcal{M}_{N,\rm{Ge}}}{\mathcal{M}_{\nu,\rm{Ge}}})\big|}
\lesssim |U^2_{\mathrm{e}4}|\lesssim \frac{\kappa_2}{m_4}\frac{1}{\big|(1- \frac{\mathcal{M}_{\nu,\rm{Xe}}}{\mathcal{M}_{N,\rm{Xe}}} \frac{\mathcal{M}_{N,\rm{Ge}}}{\mathcal{M}_{\nu,\rm{Ge}}})\big|}
\label{boundU}. 
\end{equation}

In   Fig.~\ref{fig3}, we show the upper bound on  the active-light sterile neutrino mixing angle 
$|U^{2}_{\mathrm{e}4}|$ corresponding to  the individual (solid lines) and combined (dashed lines) limits of 
half-life for $^{76}\rm{Ge}$ from GERDA \cite{gerda}. 
The area in the $|U_{\mathrm{e}4}|-m_4$ plane, that is  above this line is excluded. 
\begin{figure}[hbt]
\includegraphics[width=0.80\textwidth, angle=0]{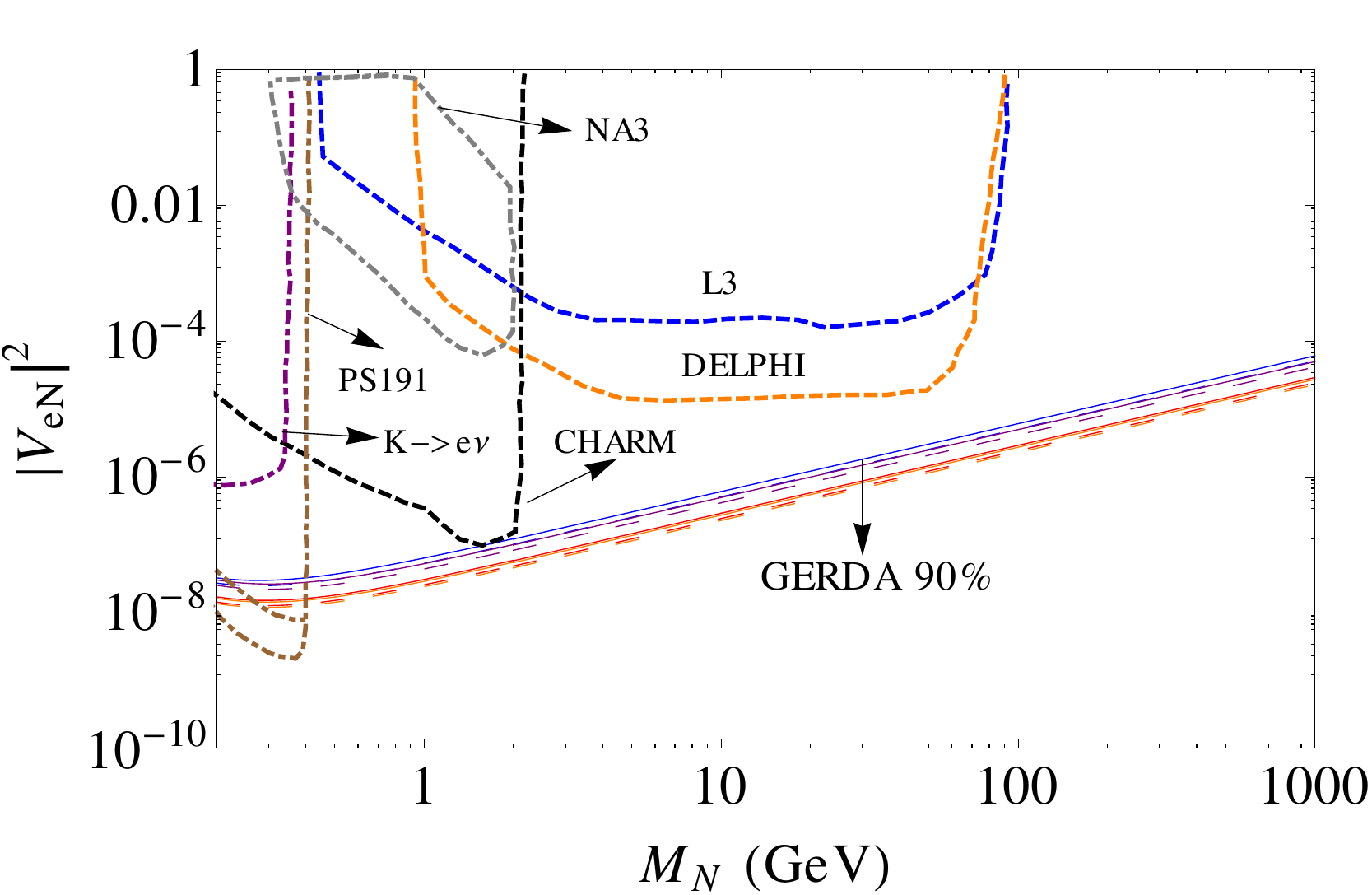}
\caption{Upper bounds  of  $|V^2_{eN}|$ that saturate the limits from  GERDA  \cite{gerda}. The different color coding  correspond to the NME uncertainty. 
In addition 
to the bound from $(\beta \beta)_{0\nu}$-decay, we also show the bounds from different other experiments, compiled in Ref.~\cite{Atre:2009rg}.} 
\label{fig4}
\end{figure}
The red, blue lines have been derived using the NMEs corresponding to the   Argonne potential (intermediate  and large size single-particle spaces, respectively) between two different nucleons and the purple and orange   lines are using the NMEs corresponding to the  CD-Bonn potential (intm and large, respectively).
For the positive claim \cite{klapdor}, 
the variation of the active sterile mixing with mass of the sterile neutrino 
is quite similar  and hence we do not show it separately.  
In this case,  the cancellation for $^{136}\rm{Xe}$ is operative mostly 
between the light sterile and heavy sterile neutrino contributions.   
For comparison we also show the other different bounds, first compiled in  
Ref.~\cite{Atre:2009rg}.  
By comparing   Fig.~\ref{fig3} with Fig.~\ref{figl}, it is evident that in the 
presence of cancellation, a larger mixing  $U_{\mathrm{e}4}$ is required to 
give the same value of the half-life. Also, as compared to Fig.~\ref{figl}, in this case the bound on active-sterile mixing angle from $\pi \to e \nu$ can compete with the bound 
from $(\beta \beta)_{0\nu}$-decay.

As we are assuming a cancellation in $^{136}\rm{Xe}$, the heavy 
sterile neutrino contribution is also constrained and a bound in the 
mass-mixing plane can be obtained. 
Using the cancellation relation  $|\eta_{l}| \mathcal{M}_{\nu, \rm{Xe}}=|\eta_N| \mathcal{M}_{N,\rm{Xe}}$ and  the values of 
$\kappa, \kappa_{1,2}$ as given in Table~\ref{tabsterile},  
the active-heavy sterile neutrino mixing angle $V_{\mathrm{e}N}$  corresponding 
 to  the GERDA and GERDA+HDM+IGEX limits \cite{gerda} is  bounded as
\begin{equation}
|V^2_{\mathrm{e}N}| \lesssim \kappa \frac{M_N}{m_\mathrm{e} m_\mathrm{p}} 
\frac{\mathcal{M}_{\nu,\rm{Xe}}}{\mathcal{M}_{N,\rm{Xe}}}\frac{1}{\big|(1- \frac{\mathcal{M}_{\nu,\rm{Xe}}}{\mathcal{M}_{N,\rm{Xe}}} \frac{\mathcal{M}_{N,\rm{Ge}}}{\mathcal{M}_{\nu,\rm{Ge}}})\big|},
\label{heavybndgerda}
\end{equation} 
while for the positive claim \cite{klapdor}, it is 
\begin{equation}
\kappa_1 \frac{M_N}{m_\mathrm{e} m_\mathrm{p}} \frac{\mathcal{M}_{\nu,\rm{Xe}}}{\mathcal{M}_{N,\rm{Xe}}}\frac{1}{\big|(1- \frac{\mathcal{M}_{\nu,\rm{Xe}}}{\mathcal{M}_{N,\rm{Xe}}} \frac{\mathcal{M}_{N,\rm{Ge}}}{\mathcal{M}_{\nu,\rm{Ge}}})\big|}\lesssim |V^2_{\mathrm{e}N}|\lesssim \kappa_2 \frac{M_N}{m_\mathrm{e} m_\mathrm{p}} 
\frac{\mathcal{M}_{\nu,\rm{Xe}}}{\mathcal{M}_{N,\rm{Xe}}}\frac{1}{\big|(1- \frac{\mathcal{M}_{\nu,\rm{Xe}}}{\mathcal{M}_{N,\rm{Xe}}} \frac{\mathcal{M}_{N,\rm{Ge}}}{\mathcal{M}_{\nu,\rm{Ge}}})\big|}.
\label{heavybnd}
\end{equation} 
Note that the  equations Eqs.~\ref{boundUgerda}, \ref{boundU}, \ref{heavybndgerda} and \ref{heavybnd}  are 
only valid for the light and heavy sterile masses smaller and larger than the exchange momentum scale, $|p| \sim 100$ MeV, respectively.
We show  the  generic equation that is valid for all mass scales in the Appendix.
Following  Eq.~\ref{heavybndgerda} and the formalism given in  Appendix, we show in  Fig.~\ref{fig4} 
 the upper bound on  the active-heavy sterile mixing $|V^{2}_{\mathrm{e} N}|$  corresponding  to  the individual and combined limits of half-life from GERDA  \cite{gerda}. The description of the different color 
coding is the same as in Fig.~\ref{fig1}. The region above the different  contours is excluded by $(\beta \beta)_{0\nu}$-decay. In this figure, for comparison  we also show the bounds coming from  other  experiments  \cite{Atre:2009rg}.  
}Again comparing Fig.~\ref{fig4} with Fig.~\ref{figh}, one can see a  
larger mixing angle $V_{\mathrm{e}N}$ required to saturate the limits on the half-life from $(\beta \beta)_{0\nu}$-decay for the case of cancellation. For the mass of the heavy sterile neutrino $M_N  \sim \mathcal{O}(100) $ MeV, the bound 
from  the beam dump experiment PS191  \cite{Bernardi:1987ek} is even stronger  than the $( \beta \beta)_{0\nu}$-decay one. In the range $M_N \sim (1-2)$ GeV, the bound from  CHARM \cite{Bergsma:1985is} can  compete with the bound from $(\beta \beta)_{0\nu}$-decay. 
 For the positive claim \cite{klapdor}, the result is similar and we do not show the corresponding region explicitly.

%\end{document}
\section{Correlation between half-lives \label{cor}} 

In this 
section, we extend our discussion of the effects of cancellations
to other isotopes. To this aim, for definiteness, we investigate how the cancellation between active and 
sterile neutrinos in $^{136}\rm{Xe}$ would influence the 
half-life of other isotopes, such as $^{100}\rm{Mo}$, $^{130}\rm{Te}$ and $^{82}\rm{Se}$ as well as $^{76}\rm{Ge}$. The ratio of half-lives in two isotopes, isotope A and isotope B, 
is  
\begin{equation}
\frac{T^{0\nu}_{1/2}({A})}{T^{0\nu}_{1/2}({B})}=\frac{G^{{B}}_{0\nu}}{G^{{A}}_{0\nu}}\frac{({\mathcal{M}_{\nu, B}}-  \frac{{\mathcal{M}_{\nu,\rm{Xe}}}}{{\mathcal{M}_{N,\rm{Xe}}}} {\mathcal{M}_{N,B}})^2}{({\mathcal{M}_{\nu,A}}-  \frac{{\mathcal{M}_{\nu,\rm{Xe}}}}{{\mathcal{M}_{N,\rm{Xe}}}} {\mathcal{M}_{N,A}})^2}.
\label{eqnratio}
\end{equation}

\begin{figure}
%\centering
\includegraphics[width=0.49\textwidth, angle=0]{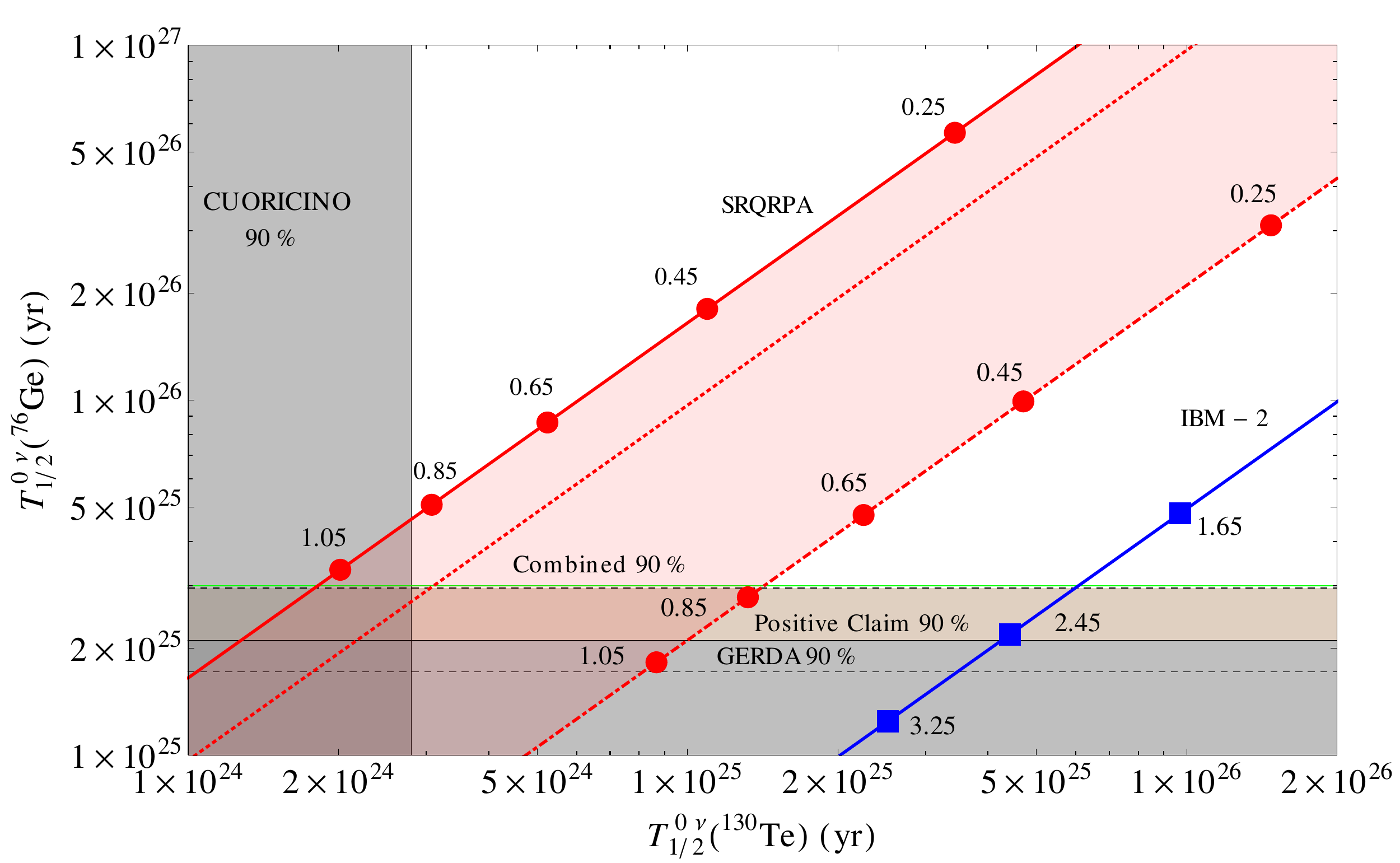}
\includegraphics[width=0.49\textwidth, angle=0]{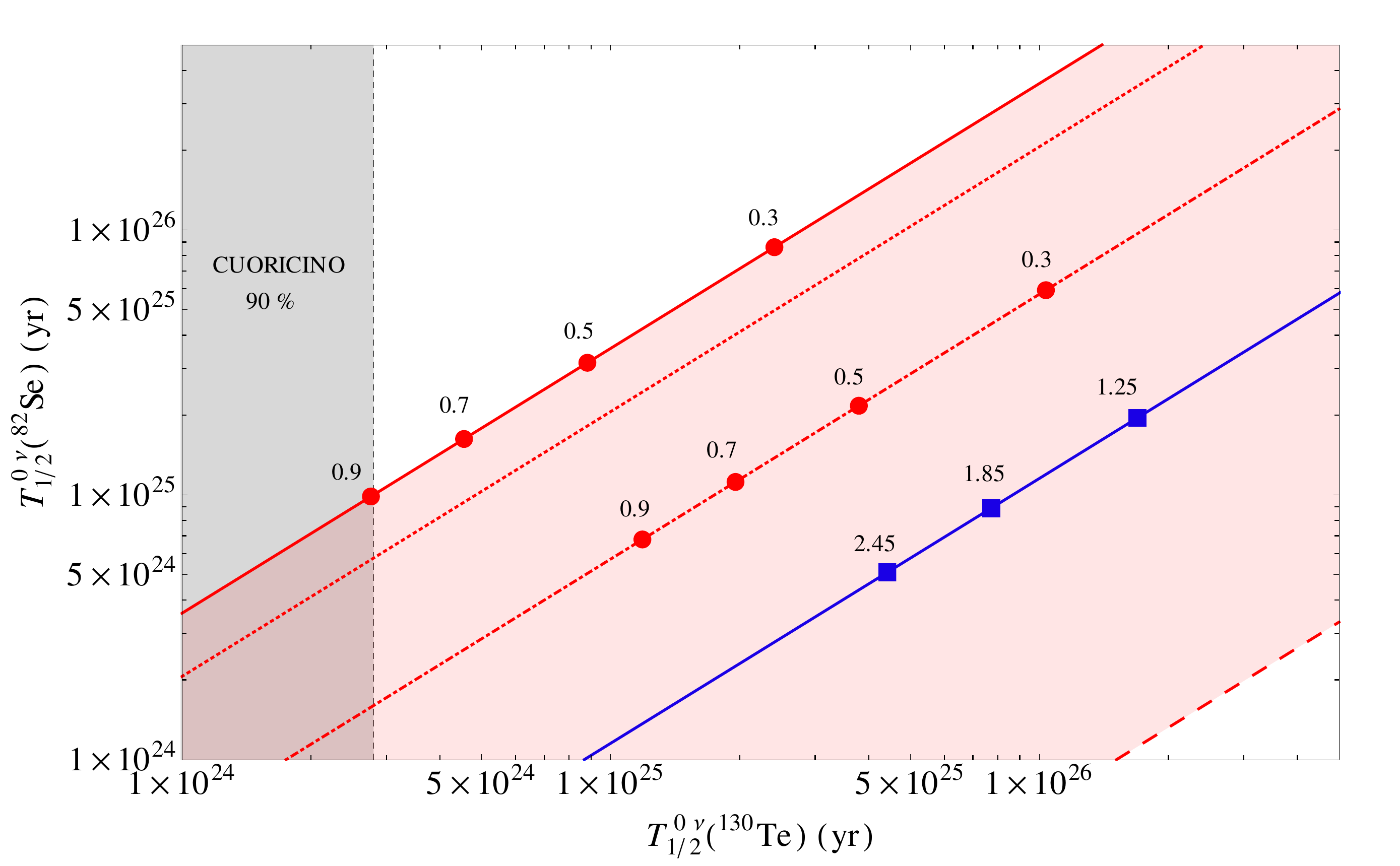}
\caption{Left Panel: 
Variation of the half-life of $^{76}$Ge with the one of $^{130}$Te, assuming a cancellation between light and heavy neutrino contributions in $^{136}\rm{Xe}$. 
The region in between the horizontal dashed black lines  corresponds to 
the positive claim   ($90\%$ C.L.) \cite{klapdor}. The black solid line correspond to the lower limit 
 of half-life of $^{76}\rm{Ge}$ from GERDA \cite{gerda}. The combined bound GERDA+HDM+IGEX 
\cite{gerda} is shown by the green horizontal line. 
 The gray shaded region is disallowed by the results from GERDA \cite{gerda} 
and  CUORICINO experiments \cite{cuo}. 
The red and blue lines correspond to the SRQRPA \cite{petcov}  and IBM-2 \cite{ibm} NME calculations, while the different numerical values represent the effective mass of light neutrino exchange $|m^{\nu}_\mathrm{ee}|$ in eV. The red dot-dashed, dashed, solid and dotted lines correspond to the NMEs that have been derived using  Argonne and CD-Bonn
potential, respectively. %The vertical gray region $T^{0\nu}_{1/2}<2.8 \times 10^{24} \, \rm{yrs}$ represent the excluded region from CUORICINO experiment  \cite{cuo}. 
Right Panel: Variation of half-life of $^{82}$Se  with the half-life of 
$^{130}$Te. The color coding is the same as for the left panel.  }
\label{fig6}
\end{figure}

\begin{figure}
\centering
\includegraphics[width=0.49\textwidth, angle=0]{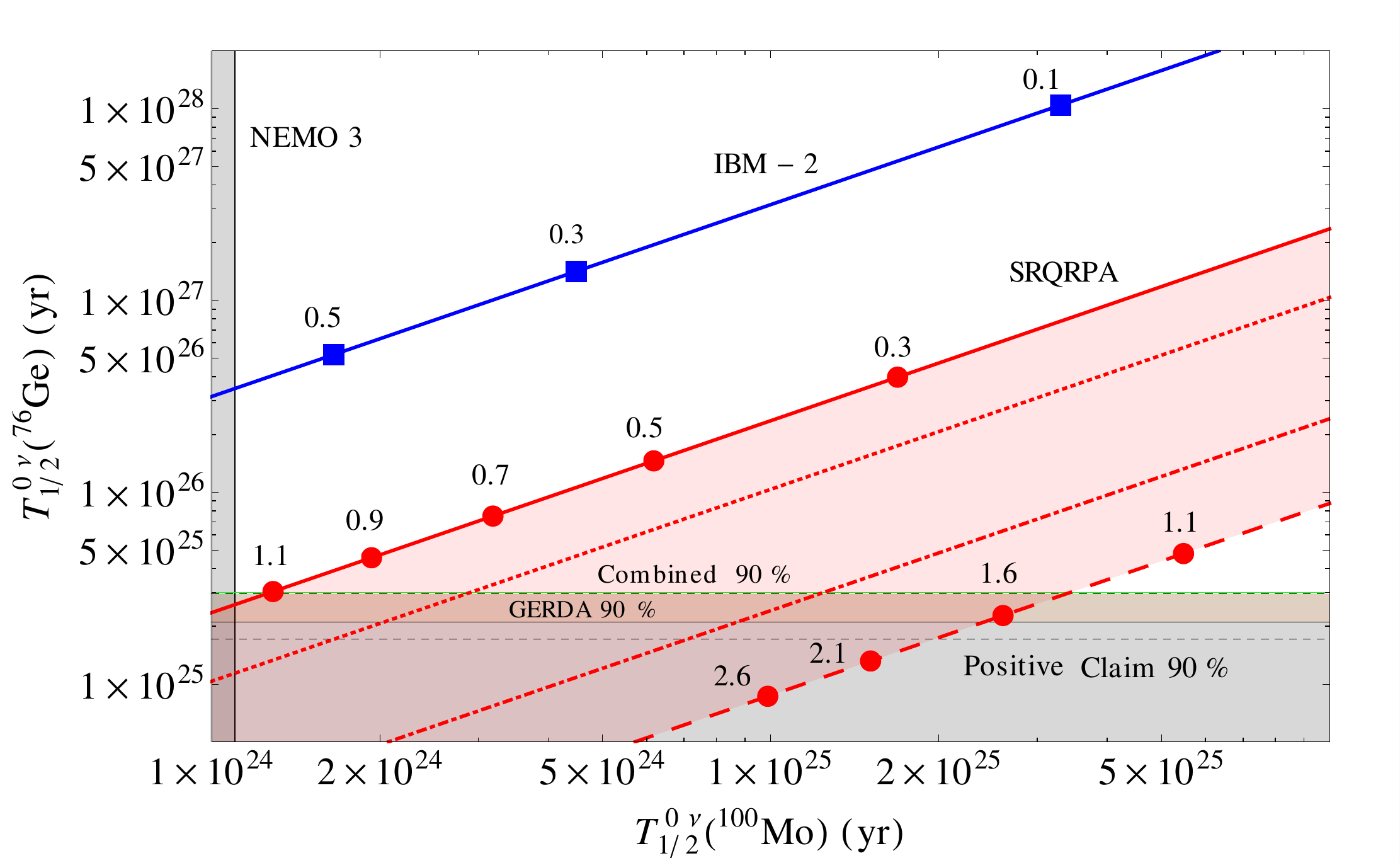}
\includegraphics[width=0.49\textwidth, angle=0]{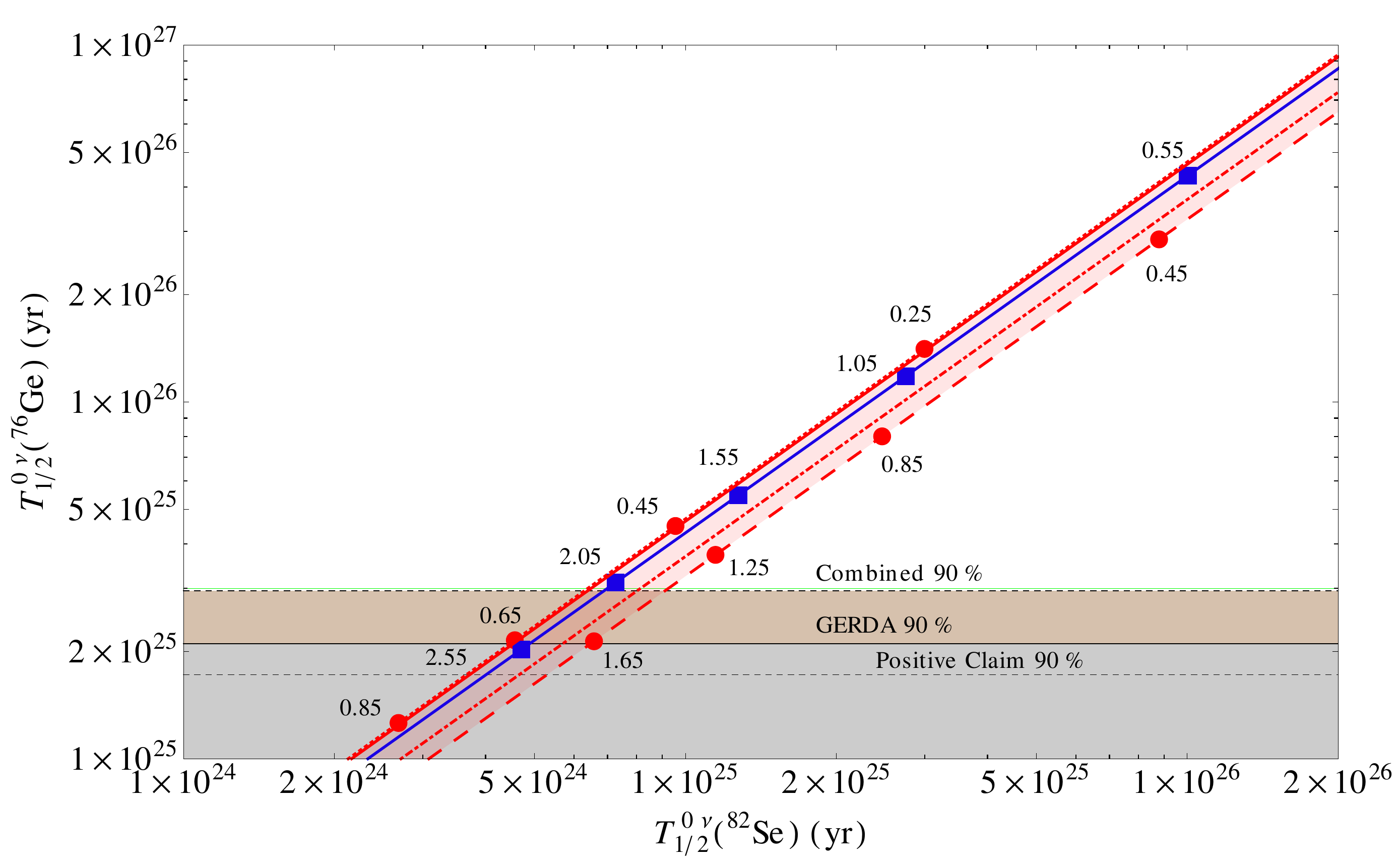}
\caption{Left Panel: Variation of the half-life of $^{76}\rm{Ge}$ with the one of $^{100}\rm{Mo}$, assuming a cancellation between light and heavy neutrino contributions in $^{136}\rm{Xe}$. The region in between the horizontal dashed black lines  corresponds to the positive  claim ($90\%$ C.L.) \cite{klapdor}. The 
black solid line correspond to the lower limit 
 of half-life of $^{76}\rm{Ge}$ from GERDA \cite{gerda} and the region below this line is excluded. The combined bound GERDA+HDM+IGEX 
\cite{gerda} is shown by the green horizontal line. 
 The red and blue lines correspond to the SRQRPA \cite{petcov} and IBM-2 \cite{ibm}
NME calculations, while 
the different numerical values represent the 
effective mass $m^{\nu}_\mathrm{ee}$ in eV. The red dot-dashed, dashed, solid and dotted lines correspond to the 
NMEs that have been derived using Argonne and CD-Bonn
potential, respectively. The vertical gray region is the excluded region from NEMO 3 \cite{nemo3} experiment. Right Panel: Variation of half-life of $^{76}\rm{Ge}$ 
with the one of $^{82}\rm{Se}$. The color coding 
is  the same as for the left panel. }
\label{fig5}
\end{figure}

Using Eq.~\ref{eqnratio}, 
 we  show the correlations between half-lives of $^{76}\rm{Ge}-^{130}${Te}, $^{82}\rm{Se}$-$^{130}\rm{Te}$, $^{76}\rm{Ge}-^{100}${Mo} and  
 $^{76}\rm{Ge}-^{82}${Se}  in Figs.~\ref{fig6} and~\ref{fig5}, respectively. 
We use different values of the NMEs which correspond to the various lines in the figures, as
specified in the captions.
The region within the two horizontal black dashed lines  correspond to positive  
claim (90$\%$ C.L.) \cite{klapdor}. The horizontal black solid line corresponds  to the individual limit from GERDA  \cite{gerda}, where the region below this line is excluded. We also show the combined GERDA+HDM+IGEX limit \cite{gerda} by the green horizontal line. 

Also in this case we can express the half-life in terms of the {\it redefined effective mass} which depends on the half-life 
${T}^{0\nu}_{1/2}(A)$  and the NMEs as
\begin{equation}
|m^\mathrm{eff}_\mathrm{ee}| = \frac{ m_\mathrm{e}}{ \sqrt{ G^A_{0\nu} {T}^{0\nu}_{1/2}(A) \mathcal{M}^2_{\nu, A}}}=\left|m^{\nu}_{\mathrm{e}\mathrm{e}} \big(1- \frac{\mathcal{M}_{\nu, \rm{Xe}}}{\mathcal{M}_{N, \rm{Xe}}}\frac{\mathcal{M}_{N, \rm{A}}}{\mathcal{M}_{\nu, \rm{A}} } \big)\right|
\label{meecor}
\end{equation}
and similarly for the isotope B. 
 The different numerical values shown in the figures represent the required effective mass $|m^{\nu}_{\mathrm{e}\mathrm{e}}|$ in eV 
for a particular 
set of half-lives $(\mathcal{T}^{0\nu}_{1/2}(A), \mathcal{T}^{0\nu}_{1/2}(B))$  of the two isotopes. Finally,  we conclude this  section by showing the individual 
prediction of half-lives of $^{130}\rm{Te}$ and $^{100}\rm{Mo}$ in Table~\ref{tab1} and 
Table~\ref{tab2}, respectively.

\begin{table}[h!]
\begin{center}
\begin{tabular}{c|c|c|c|c|c|c|c|c}\hline\hline
\multicolumn{6}{c|}{NME} &  \multicolumn{3}{c}{$T_{1/2}^{0\nu}(^{130}{\rm Te})$ ($10^{25}$ yr)}\\ \cline{1-9}
 ${\cal M}_{0\nu}(^{76}{\rm Ge})$ & ${\cal M}_{N}(^{76}{\rm Ge})$ & ${\cal M}_{0\nu}(^{136}{\rm Xe})$ & 
${\cal M}_{N}(^{136}{\rm Xe})$ & ${\cal M}_{0\nu}(^{130}{\rm Te})$ & ${\cal M}_{N}(^{130}{\rm Te})$ & Positive claim  & GERDA & Combined 
 \\ \hline
4.75 & 232.8 & 2.29 & 163.5 & 4.16 & 234.1 & 0.82-1.40 & 0.997  & 1.42 \\ 
5.44 &  264.9 &  2.75 &  159.7 &  4.18 &  239.7  & 80.2-137.97 & 98.0 & 140.0 \\ 
5.11 &  351.1 & 2.95 &  166.7 & 4.62 &  364.3 & 0.10-0.18 & 0.13 & 0.18 \\ 
5.82 &  411.5 &   3.36 &  172.1 & 4.70 &  384.5 & 0.18-0.31 & 0.22 & 0.31 \\
\hline \hline
\end{tabular}
\end{center}
\caption{Predictions of the half-life  $T_{1/2}^{0\nu}(^{130}{\rm Te})$ that corresponds to the 
i) positive claim  in $^{76}$Ge: $T^{0\nu}_{1/2}=2.23^{+0.73}_{-0.51}
\times 10^{25}$ ( $90\%$ C.L.) \cite{klapdor}; ii) saturates the GERDA and iii) the  GERDA+HDM+IGEX  (Combined) limits 
 $T^{0\nu}_{1/2}> (2.1, 3.0) \times 10^{25} \rm{yrs}$ of half-life \cite{gerda}, while satisfying the limit of half-life from EXO-200, KamLAND-Zen \cite{Auger:2012ar, Gando:2012zm} as an artifact of cancellation between light and heavy states.  We have 
used the NMEs from \cite{petcov}. Following \cite{Kotila:2012zza}, the phase 
space factors that have been used are
$G_{0\nu}(^{76} \rm{Ge})=5.77 \times 10^{-15} \rm{yrs}$, $G_{0\nu}(^{136} \rm{Xe})=3.56 \times 10^{-14} \rm{yrs}$ 
and $G_{0\nu}(^{130} \rm{Te})=3.47 \times 10^{-14} \rm{yrs}$. 
}
\label{tab1}
\end{table}

\begin{table}[h!]
\begin{center}
\begin{tabular}{c|c|c|c|c|c|c|c|c}\hline\hline
\multicolumn{6}{c|}{NME} &  \multicolumn{3}{c}{$T_{1/2}^{0\nu}(^{100}{\rm Mo})$ ($10^{25}$ yr)}\\ \cline{1-9}
${\cal M}_{0\nu}(^{76}{\rm Ge})$ & ${\cal M}_{N}(^{76}{\rm Ge})$ & ${\cal M}_{0\nu}(^{136}{\rm Xe})$ & 
${\cal M}_{N}(^{136}{\rm Xe})$ & ${\cal M}_{0\nu}(^{100}{\rm Mo})$ & ${\cal M}_{N}(^{100}{\rm Mo})$& Positive claim  & GERDA & Combined\\ \hline
4.75 & 232.8 & 2.29 & 163.5 & 4.39 & 249.8 & 0.71-1.23 & 0.87 & 1.24 \\ 
5.44 &  264.9 &  2.75 &  159.7 &  4.79 &  259.7  & 1.95-3.35 & 2.38 & 3.40 \\ 
5.11 &  351.1 & 2.95 &  166.7 & 4.81 &  388.4 & 0.07-0.13 & 0.09 & 0.13 \\ 
5.82 &  411.5 &   3.36 &  172.1 & 5.15 &  404.3 & 0.17-0.29 & 0.20 & 0.29 \\
\hline \hline
\end{tabular}
\end{center}
\caption{The same as Table~\protect\ref{tab1} but for  $^{100}{\rm Mo}$.
  Following \cite{Kotila:2012zza}, the phase 
space factors that we have  used are
$G_{0\nu}(^{76} \rm{Ge})=5.77 \times 10^{-15} \rm{yrs}$, $G_{0\nu}(^{136} \rm{Xe})=3.56 \times 10^{-14} \rm{yrs}$ 
and $G_{0\nu}(^{100} \rm{Mo})=3.89 \times 10^{-14} \rm{yrs}$. 
}
\label{tab2}
\end{table}

\section{Model-Seesaw realizations \label{mod}}

As discussed above, the cancellation between light and heavy contributions to neutrinoless double beta decay in one isotope requires very specific values of neutrino masses and mixing angles. 
In this section we discuss how such values can emerge from theoretical models. The  most natural framework embedding sterile neutrinos is the Type-I seesaw  mechanism. Typically, heavy sterile neutrinos are introduced at or just below the GUT scale leading to light neutrino masses.
If their mass is larger than few tens of TeV, the contribution in $(\beta \beta)_{0\nu}$-decay  would be negligibly small~\cite{ibarra, MSV, Pascoli}. However, sterile neutrino can have much smaller masses, even well below the electroweak scale, e.g. in low energy see-saw models \cite{lowscaleseesaw, lowscaleseesaw1}. A lot of attention has been recently devoted to sterile neutrino states with masses lighter than TeV scale  in $(\beta \beta)_{0\nu}$-decay in
Refs.~\cite{ibarra,MSV,Pascoli,blennow,meroni2013}. Below we discuss 
specific models which can accommodate light as well as  heavy sterile neutrinos and lead to the cancellations we are interested in. 

%\end{document}
\subsection{Model A - Light Active and Heavy Sterile Neutrinos}

We consider first the case in which all sterile neutrinos are heavy, having masses larger than the momentum exchange scale $|p| \sim 100$ MeV, see Section~\ref{heavy}. 
We consider $n$ generations of sterile neutrinos $(\hat{N}_i, \hat{N'}_i)$ denoted in the flavor basis. 
In the $(\nu, \hat{N}, \hat{N'})$  basis, 
the mass matrix of active+sterile neutrinos has the following form 
\be
M_n=\pmatrix {0 & \alpha^T & m^T_D \cr \alpha & \mu  & m^T_S \cr m_D & m_S & m_R} ~,
\label{eq:extended}
\ee
where $\mu$ and $m_R$ are   two lepton number violating parameters~\footnote{Depending on the choice of the lepton number assignment for the $\hat{N}, \hat{N'}$ fields, different parameters in the mass matrix will be lepton number violating. Here, we adopt a common choice in which $m_D, m_S$ and $m_R$ are large masses and $\mu$ is very small.}.
Particularly interesting phenomenology
arises for the hierarchy $m_R>m_S>m_D \gg \mu, \alpha$ and $\frac{m^2_S}{m_R}\gg \mu,\alpha$ which
will lead to the Extended seesaw scenario \cite{Kang-Kim, Parida}. 
 We denote the mass basis as $(\nu_m, N, {N'})$. The mass of the 
sterile neutrinos $N$, ${N'}$ are  obtained by  diagonalizing 
Eq.~\ref{eq:extended}  and 
are given by 
\begin{eqnarray}
{m}_N & \simeq & -m^T_S m^{-1}_R m_S,   \\
{m}_{{N'}} & \simeq &  m_R.
\end{eqnarray}
Let us note that for simplicity we call sterile neutrinos both the flavor states and the massive states which are mainly in the sterile neutrino direction.
From the inequality $m_R> m_S$ it follows that $m_{{N'}}>m_N$. 
In the following discussion we consider the simplest case in which $\alpha$ is negligibly small. The mass matrix of the active neutrino depends on the small lepton number violating parameter $\mu$ and is 
\begin{equation}
m_{\nu} \simeq  m^T_D (m^T_S)^{-1} \mu (m_S)^{-1} m_D.
\end{equation}
%%%%
Depending on the the values of $\frac{m_D}{m_S}$ and $\mu$, light neutrino of eV  mass
can be obtained. The mixings of sterile neutrinos $N$ and ${N'}$ with active neutrinos are 
\begin{eqnarray}
U_{\mathrm{e} N} &\simeq &  (m^{\dagger}_D(m^{-1}_S)^{\dagger})_{\mathrm{e} N} , \\ 
U_{ \mathrm{e}{N'}} & \simeq &  (m^{\dagger}_D m^{-1}_R)_{ \mathrm{e}{N'}}. 
\end{eqnarray}
Note that, while the light neutrino mass depends on the lepton number violating parameter $\mu$, the active-sterile neutrino mixing is independent of this parameter to leading order. Hence,  in this case, one can have large active-sterile neutrino mixing  while neutrino masses are kept small thanks to the $\mu$ parameter.

This seesaw scenario has been explored previously in \cite{Kang-Kim, Parida}.
In this work we are interested to study if the heavy neutrinos in this model can satisfy the cancellation conditions and give large contributions in $(\beta \beta)_{0\nu}$-decay, thanks to not-too large masses and large mixings \cite{MSV}. 
 For simplicity, we drop the active and sterile neutrino indexes and focus on the order of magnitude of the parameters of the mass matrix. Interesting flavor effects could be present but they are beyond the scope of the present analysis.
For heavy sterile neutrinos of masses $m_N, m_{{N'}}\gg 100 $~MeV, the  amplitudes corresponding to the $N$ and ${N'}$ contributions to $(\beta \beta)_{0\nu}$-decay  are  \cite{MSV, Pascoli}
\begin{eqnarray}
\left|\mathcal{A}_N\right|=\left|\frac{{U}^2_{\mathrm{e}N}}{m_N}\right|  \sim \left|\frac{m^2_D}{m^2_Sm_N}\right|, \\
\left|\mathcal{A}_{{N'}}\right| = \left|\frac{{U}^2_{\mathrm{e}N'}}{m_{{N'}}}\right|\sim \left|\frac{m^2_D}{m^2_Rm_{N'}}\right|.
\label{ampste}
\end{eqnarray}
As  $m_R\gg m_S$, the role of ${N'}$ is suppressed both by the large mass and small mixing and we neglect it with respect to $N$ in the following discussion.

Including the contributions from light neutrinos and  the sterile state $N$, 
the half-life of neutrinoless double beta decay for a particular isotope A  is given by
\be
\frac{1}{T^{0\nu}_{1/2}(A)} \approx G^A_{0\nu}\Big(\frac{|m^{\nu}_\mathrm{ee}|^2}{m^2_\mathrm{e}}\mathcal{M}^2_{\nu, A}+m^2_\mathrm{p} \left|\frac{m^2_D}{m^2_Sm_N} \right|^2 \mathcal{M}^2_{N,A}+
2 \cos \alpha \, 
\left|\frac{m^{\nu}_\mathrm{ee}}{m_\mathrm{e}} \right| \,\left|\frac{m^2_Dm_\mathrm{p}}{m^2_Sm_N} \right| \, \mathcal{M}_{\nu,A}\, \mathcal{M}_{N,A}\Big).
\label{hl}
\ee

Again, for definiteness, we consider the case in which a cancellation is operative for $^{136}\rm{Xe}$ isotope and find its implication for $^{76}\rm{Ge}$. Using  the cancellation condition $|\eta_{\nu}|\mathcal{M}_{\nu,\rm{Xe}}=|\eta_N| 
\mathcal{M}_{N, \rm{Xe}}$, 
we get the following relation between different parameters  
\begin{equation}
\frac{\mu}{m_\mathrm{e}} \left(\frac{m_D}{m_S}\right)^2 \mathcal{M}_{\nu,\rm{Xe}}=\left(\frac{m_D}{m_S} \right)^2 \frac{m_\mathrm{p}m_R}{m^2_S}\mathcal{M}_{N,\rm{Xe}}, 
\end{equation}
which simplifies to 
\begin{equation}
\mu =\left(\frac{m_R}{m_S^2}\right)\left(\frac{\mathcal{M}_{N,\rm{Xe}}}{\mathcal{M}_{\nu,{\rm{Xe}}}}\right) m_\mathrm{e}m_\mathrm{p}~.
\label{cance-ex}
\end{equation}
Using the definition $ p^2_{\rm{Xe}} \equiv -m_\mathrm{e} m_\mathrm{p} (\frac{\mathcal{M}_{N,{\rm{Xe}}}}{\mathcal{M}_{\nu,{\rm{Xe}}}}) $, we get 
$\mu=(\frac{m_R}{m^2_S})|p^2_{\rm{Xe}}| \simeq |p^2_{\rm{Xe}}|/m_N $.  {{We recall that $m_N \gg \sqrt{|p^2_{\rm{Xe}}|}$, implying that $\mu \ll   m_N$ in agreement with the original assumption of the hierarchy of the neutrino mass parameters}. Taking the typical range for $m_N$ given by $100 \ \mathrm{MeV} $--$ 10^6 \ \mathrm{GeV}$, we find that the $\mu$ parameter will be typically small, $\mu \sim 0.1 $--$ 10^{-8}$~GeV, as originally assumed. 

Using the above  \ref{hl} and Eqs.~\ref{cance-ex}, we can express the lepton number violating parameter $\mu$ as a function of the half-life time of $^{76}\rm{Ge}$
\be
\mu=\frac{m^2_S}{ m^2_D} \frac{m_{\mathrm{e}}}{\sqrt{G^{\rm{Ge}}_{0\nu}T^{0\nu}_{1/2}(^{76}\rm{Ge})}}\frac{1}{{{\big({\mathcal{M}}_{\nu,\rm{Ge}}}-  \frac{{\mathcal{M}_{\nu,\rm{Xe}}}}{{\mathcal{M}_{N,\rm{Xe}}}} {\mathcal{M}_{N,\rm{Ge}}\big)}}}. 
\label{extkl}
\ee

\begin{figure}
%\centering
\includegraphics[width=0.495\textwidth, angle=0]{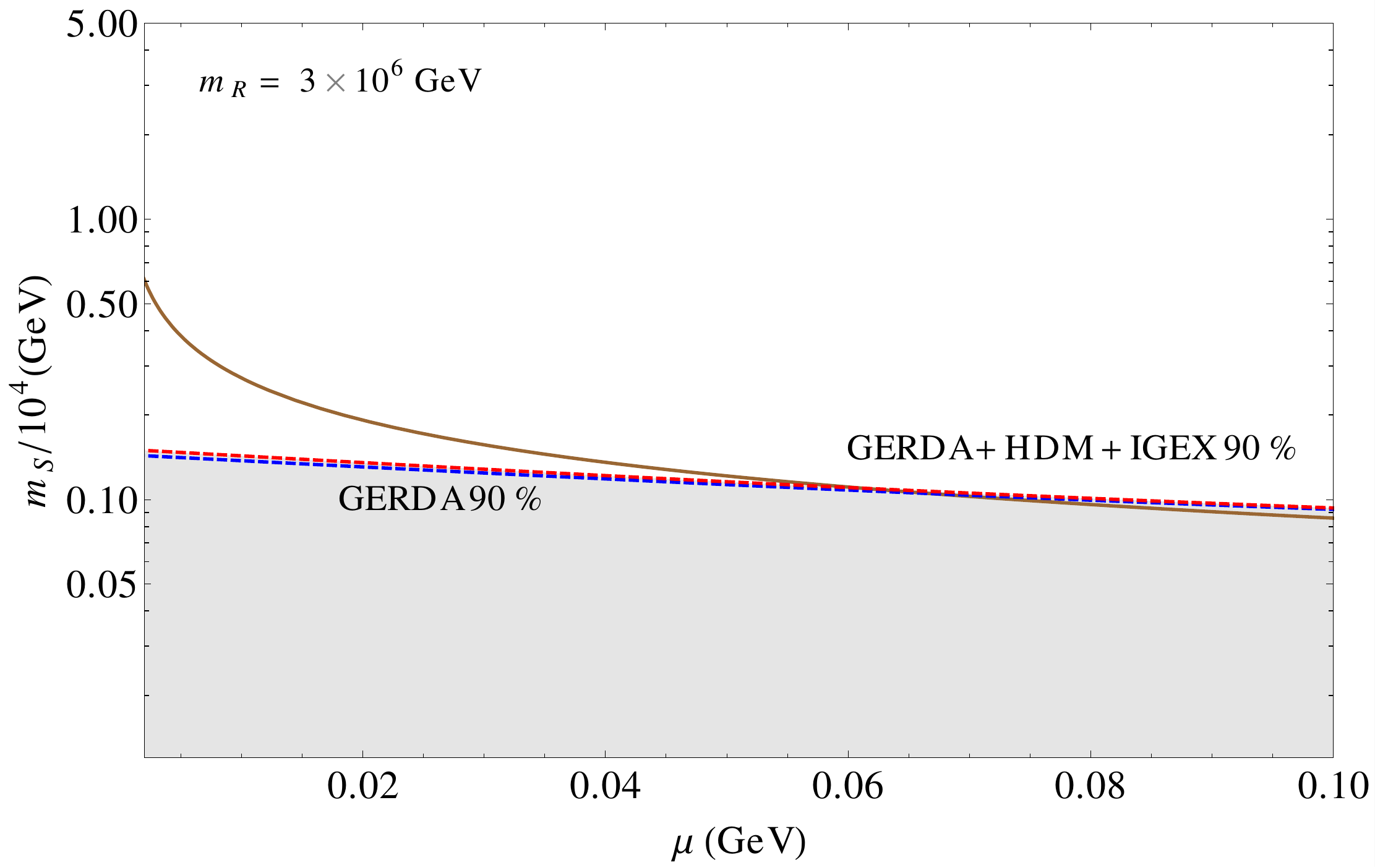}
\includegraphics[width=0.495\textwidth, angle=0]{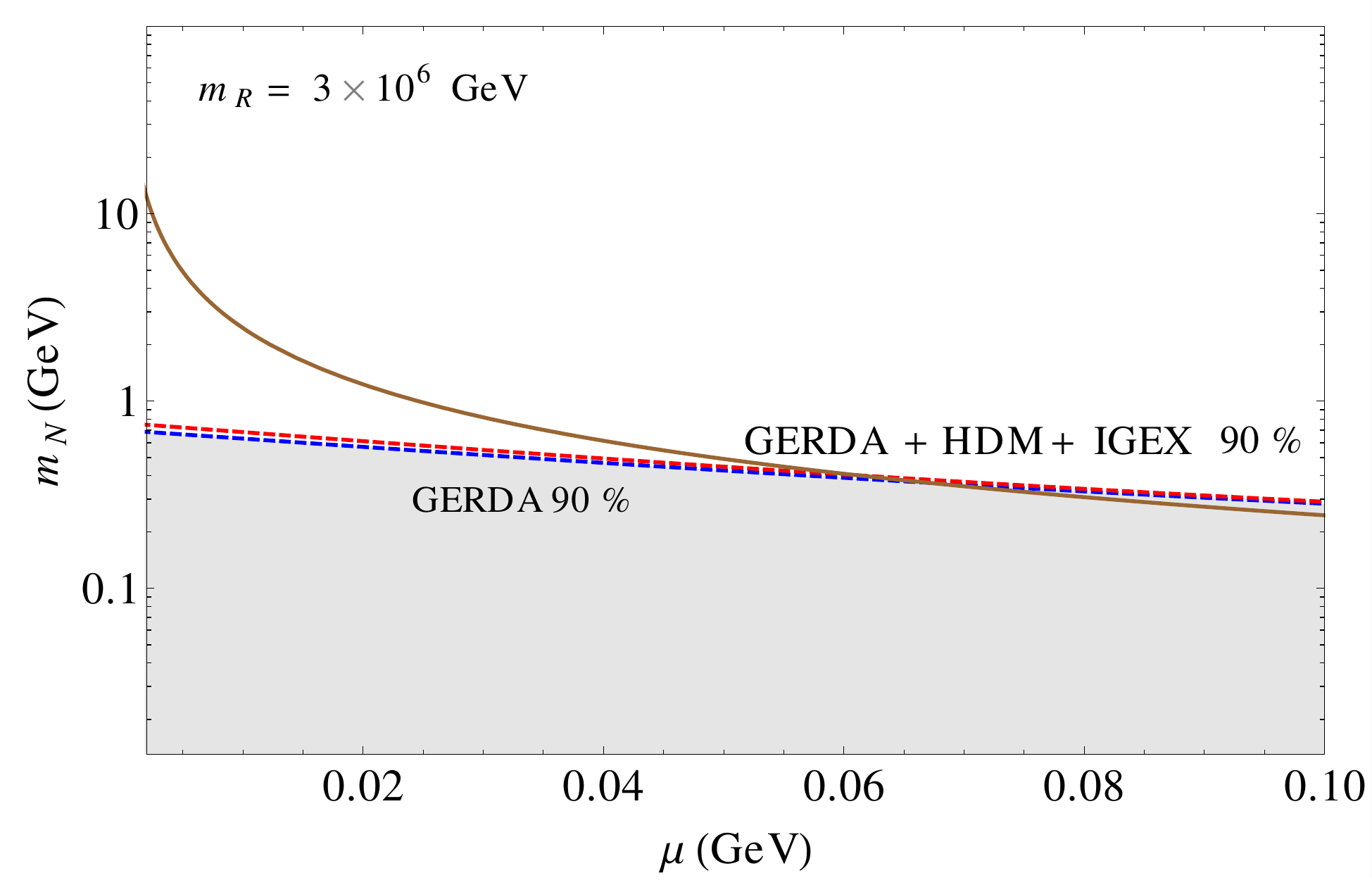}
\caption{
Variation of  $m_S$ and the mass $m_N$ of the sterile neutrino $N$ vs $\mu$ for $m_R=3 \times 10^6$ GeV and $m_D=0.1$ GeV. The 
gray shaded region is disallowed from GERDA experiment. The brown line corresponds to the cancellation condition between the light active and heavy sterile neutrinos. The red and blue lines correspond to the  half-life of 
GERDA and GERDA+HDM+IGEX limits  \cite{gerda}.}
\label{fig7}
\end{figure}

Below we discuss the different constraints on the parameters of the model, 
that satisfy the cancellation in $^{136}\rm{Xe}$ and the lower limit of 
half-life from GERDA \cite{gerda}. 
In the left panel of Fig.~\ref{fig7} the blue and red lines represent the variation of $m_S$ with the lepton number 
violating parameter $\mu$  for  representative values of the parameters, $m_R=3 \times 10^6$ GeV,  $m_D=0.1 \, \rm{GeV}$, 
that saturate the individual and combined limits from GERDA \cite{gerda}.
We have used the NMEs  
$\mathcal{M}_{\nu}(^{76}\rm{Ge})=5.82$ and $\mathcal{M}_{N}(^{76}\rm{Ge})=411.5$ \cite{petcov} that correspond  to the CD-Bonn potential between two nucleons. The gray region below this line is   not allowed.  
The brown line corresponds 
to Eq.~\ref{cance-ex} and satisfies 
the cancellation in $^{136}\rm{Xe}$ for the NMEs $\mathcal{M}_{\nu}(^{136}\rm{Xe})=3.36$ and 
$\mathcal{M}_{N}(^{136}\rm{Xe})=172.1$ \cite{petcov}. 
For the positive claim \cite{klapdor}, the variation of $m_S$ is similar and we do not show it explicitly. In the right panel of Fig.~\ref{fig7} 
we  show the corresponding 
variation of the physical mass $m_N$ of the sterile neutrino $N$. 
The intersection of the blue, red line  and brown line 
represents the point in  the parameter space where the active neutrino,  together with the heavy sterile neutrinos can simultaneously 
 saturate the bounds from GERDA and GERDA+HDM+IGEX \cite{gerda}, as well as the bound from EXO-200, KamLAND-Zen and the combined one \cite{Auger:2012ar, Gando:2012zm}. The light neutrino mass in the intersection region is $m_{\nu} \sim 0.66 $ eV.

In principle one can also consider the limit $\mu \to 0$. Although light neutrinos would be massless at tree level, 
a nonzero mass is  generated at loop level. The finite one loop correction to the light neutrino mass for this model has been computed in Ref.~\cite{Pascoli}
{\bea
\label{loopmass}
\delta m_{\nu}
&=&\frac{1}{(4 \pi v)^2}\frac{m^2_D}{2} \left\{
\left( \frac{3{m_N}\ln\left({m_N}^2/M^2_Z\right)}{{m_N}^2/M^2_Z-1}+\frac{{m_N}\ln\left({m_N}^2/M^2_H\right)}
{{m_N}^2/M^2_H-1}\right)\cos^2\theta\right.
\nonumber\\
&+&
\left.\left(\frac{3{m}_{N'}\ln\left({m}^2_{N'}/M^2_Z\right)}{{m}^2_{N'}/M^2_Z-1}+\frac{{m}_{N'}\ln\left({m}^2_{N'}/M^2_H\right)}
{{m}^2_{N'}/M^2_H-1}\right)\sin^2\theta \right\}, 
\eea}
where  $\theta$ is the mixing angle between the sterile states $N$ and ${N'}$,

\begin{equation}
\tan\theta=\frac{m_R-\mu + \sqrt{4m^2_S+(m_R-\mu)^2}}{2m_S}.
\label{mixangle}
\end{equation}
For $\mu=0$, the cancellation between light active and heavy sterile state
leads to the following relation
\bea
\delta m_{\nu} \frac{\mathcal{M}_{\nu,\rm{Xe}}}{m_\mathrm{e}}=\frac{m^2_D}{m^2_S} \frac{m_\mathrm{p}}{m_N} \mathcal{M}_{N,\rm{Xe}} ~.
\label{canceloop}
\eea

 Using Eq.~\ref{lh}, and the NMEs given above, $\delta m_{\nu} \sim 0.66 \, \rm{eV}$ is required  to satisfy the individual limit from GERDA  \cite{gerda}. 
As for the previous case, it is possible to identify the range of parameters 
that satisfy the cancellation condition and the constraints from GERDA \cite{gerda}. Here, we present  a simple numerical example:  $m_R=10^8 \, \rm{GeV}$, while  $m_D=0.75 \, \rm{GeV}$, 
$m_S=6.73 \times 10^3\, \rm{GeV}$. In this case, the mass of the two sterile neutrinos are 
 $m_N \sim 0.45 \, \rm{GeV}$  and $m_{{N'}} \sim 10^8 \, \rm{GeV}$.  For this choice of 
parameters, both the cancellation condition Eq.~\ref{canceloop}  
and the required value of light neutrino mass $\delta m_{\nu}=0.66 $ eV to 
saturate the limit from GERDA \cite{gerda} can be achieved. Similar discussion holds for the positive claim \cite{klapdor}.

\subsection{Model B---Light and Heavy Sterile Neutrino}

We discuss  the simplest seesaw realization  which can accommodate one light and one
 heavy sterile neutrinos and the cancellation in $(\beta \beta)_{0\nu}$-decay, 
corresponding to the discussion of Section~\ref{light-heavy}. This can be achieved using the mass matrix presented 
in Eq.~\ref{eq:extended}} of the previous section with the addition of a Type-II seesaw mass term of light neutrinos. For 
simplicity we consider that the Majorana mass matrix of the two sterile neutrinos is diagonal. We denote the sterile neutrinos
as $\tilde{N}$ and $\tilde{N'}$ in this  basis.   
The neutrino mass matrix in this basis  is 
\be
M=\pmatrix{m_{\Delta} & \tilde{\alpha} & \tilde{m}_D \cr 
\tilde{\alpha} & {\mu} & 0 \cr
\tilde{m}_D & 0 & {m}_R}.
\ee
For the sterile neutrino masses $m_{R},\mu> \tilde{\alpha}, \tilde{m}_D, m_{\Delta}$, the light neutrino mass term and its mixing with the sterile neutrino are
\be
m_{\nu}=m_{\Delta}-\frac{\tilde{\alpha}^2}{\mu}-\frac{\tilde{m}^2_D}{m_R}, \\
\label{massm}
\ee
and 
\be
\nu \sim \nu_m+\frac{\tilde{\alpha}}{\mu} N+\frac{\tilde{m}_D}{m_R}{N'}.
\label{mix}
\ee 
The other two sterile neutrino masses are $\mu$ and $m_R$, respectively.
For $\mu$ and $m_R$ to be  smaller and larger than the momentum exchange scale, the total contribution in 
$(\beta \beta)_{0\nu}$-decay  is
\be
\frac{1}{T^{0\nu}_{1/2}}=G_{0 \nu}\left| (m_{\nu}+ \frac{\tilde{\alpha}^2}{\mu^2} \mu)\frac{\mathcal{M}_{\nu}}{m_\mathrm{e}}-\frac{\tilde{m}^2_D}{m^2_R}\frac{m_\mathrm{p}}{m_R}\mathcal{M}_N\right|^2.
\label{cont}
\ee
To discuss the simplified constraints on the parameter space, we consider the case 
where $m_{\nu}\ll \frac{\tilde{\alpha}^2}{\mu}$ which requires additional fine tuning 
between the different terms in Eq.~\ref{massm}. If this is the case, then 
 the cancellation between the light sterile 
and heavy sterile contribution for $^{136}\rm{Xe}$ gives the following condition
\be
\tilde{m}^2_D=\tilde{\alpha}^2 \frac{m^3_R}{\mu} (\frac{M_{\nu,\rm{Xe}}}{M_{N,\rm{Xe}}})\frac{1}{m_\mathrm{p} m_\mathrm{e}}.
\label{cond1}
\ee
In addition, we consider that the light and  heavy sterile contribution satisfy the bounds for $^{76}\rm{Ge}$ or the positive claim \cite{klapdor}. Using Eq.~\ref{cont}, we get 
\be
\tilde{\alpha}^2=\frac{\mu \, m_\mathrm{e}}{\sqrt{({\mathcal{M}_{\nu,\rm{Ge} }}-  \frac{{\mathcal{M}_{\nu,\rm{Xe}}}}{{\mathcal{M}_{N,\rm{Xe}}}} {\mathcal{M}_{N,\rm{Ge}}})^2}} \frac{1}{\sqrt{{T}^{0\nu}_{1/2}(\rm{^{76}Ge}) G^{\rm{Ge}}_{0\nu} }}.
\label{cond2}
\ee
In this case,  the bound from $^{136}\rm{Xe}$ is possible to escape. 
As an example we consider the sterile neutrino masses $\mu = 0.01 \, \rm{GeV}$ and $m_R=1\, \rm{GeV}$ and the different nuclear matrix elements given in the previous section. For the choice of parameters $\tilde{\alpha}=2.54 \times 10^{-6}\, \rm{GeV}  $ and $\tilde{m}_D=1.61 \times 10^{-4}\, \rm{GeV} $, the positive claim \cite{klapdor} as well as the cancellation condition  are possible to satisfied.  In this case, the sterile neutrino contribution to the neutrino mass matrix would be $26.85\, \rm{ eV}$. Hence, the amount of fine tuning that is
required in this case to obtain $m_{\nu} \sim 0.1\, \rm{eV}$ is of similar order, which can be achieved by adjusting $m_{\Delta}$.
Note that, the contribution  from Higgs triplet in $(\beta \beta)_{0\nu}$-decay depends on the Higgs triplet mass. For very large mass of Higgs triplet, their
contribution will be negligibly small and can safely be avoided.

\section{Conclusion \label{conclu}} 

In light of  recent experimental results, in this work we  have carefully analyzed  the effect of interference  in $(\beta \beta)_{0\nu}$-decay. Most studies assume that the light Majorana neutrino exchange is the dominant mechanism mediating this process. However, any beyond-the-standard-model framework, which is required to generate light Majorana neutrino masses, will also induce neutrinoless double beta decay directly due to its lepton number violating parameters and could give a relevant contribution. If the  different contributions are  sizable,  they can  interfere  either  constructively or destructively. For definiteness, we consider the case of  heavy sterile neutrinos with  masses larger than the momentum exchange, $|p| \sim 100$ MeV, and light sterile neutrinos. If their masses are smaller than  TeV scale and  if their mixings 
with the electron neutrinos are sizable, 
they can saturate the current bounds of half-life \cite{MSV}. 

If a complete cancellation is at work, the half-life of $^{136}\rm{Xe}$ is infinite and any constraint on it   would be automatically satisfied, independently from the results for other isotopes. Due to the different nuclear matrix elements, 
only a partial interference will be present for other nuclei. As an example, motivated by the not-yet-completely-excluded claim
of $(\beta \beta)_{0\nu}$-decay in $^{76}\rm{Ge}$, we have studied the predictions in detail for the half-life in 
this isotope and  the correlations with other nuclei. 

 A large value of the effective mass $m^{\nu}_{\mathrm{ee}} \sim \mathcal{O}(0.1)$ eV is required to satisfy the 
positive claim \cite{klapdor} or to saturate the current bounds from $(\beta \beta)_{0\nu}$-decay experiments. 
For the case in which   only three light active neutrinos are present,  their masses are required to be in the 
  quasidegenerate regime. However, this possibility is strongly constrained by the stringent bounds from 
cosmology. If the cancellation between light and heavy neutrino exchange is at work, 
 the redefined effective Majorana mass gets suppressed. Or in other words, a larger value 
of the true effective mass $m^{\nu}_{\mathrm{ee}} $ (0.66 eV-1.67 eV will saturate GERDA for SRQRPA calculation) is required to have the same half-life.  Hence, 
bigger values
of neutrino masses are needed. {As a result}, if the {\it 
redefined effective mass}  saturates the limits from $(\beta \beta)_{0\nu}$-decay, the 
tension with cosmological data becomes even more severe. 
In the next few years, quasi-degenerate values of neutrino masses will be 
tested by the $\beta$-decay experiment KATRIN \cite{katrin}, providing additional constraints on this 
possibility.

The tension with cosmology can be weakened  if we also consider light sterile neutrinos. Depending on the mass and mixing, 
 light sterile neutrinos can give a large  contribution in $(\beta \beta)_{0\nu}$-decay, and can even saturate current 
 limits. On the other hand, the bounds from cosmology are 
 relevant only if their masses are in the  eV range for 
the values of mixing angles of interest. 
Neutrinoless double beta decay turns out to be the most sensitive probe of 
these {sterile} neutrinos. 
For masses in the range 10 eV-100 KeV, 
the bounds from  beta-decay experiments are weaker than that of $(\beta \beta)_{0\nu}$-decay by a factor of 
$U^2_{\mathrm{e}4} \sim \mathcal{O}(10-100)$. 
For sterile neutrinos of  $ \mathcal{O}(100)$ MeV masses, the constraints from the 
peak search in  $\pi \to e \nu$  and the beam dump experiment PS191 reach   similar sensitivity 
as  $(\beta \beta)_{0\nu}$-decay. 
 In the presence of cancellations,  a 
larger value of  active-sterile mixing angle is required to obtain the same value of half-life. 
Hence, the bounds from  experiments,  such as  beta decays, $\pi \to e\nu$,   PS191 and even CHARM  become competitive with $(\beta \beta)_{0\nu}$-decay, making it easier to test the parameters required for a cancellation.

A direct test of destructive interference, being at work in a certain nuclei, will be given by the 
measurement of the half-life in several isotopes \cite{petcov, Faessler:2011qw, Faessler:2011rv, f2010}. In the case under study, 
the cancellation between light active/sterile and heavy sterile neutrino exchange  in $^{136}\rm{Xe}$ 
will lead to a definite prediction of the half-lives of  other isotopes. 
If we take {$m^{\nu}_{\mathrm{ee}} \sim (0.5-1) \, \rm{eV}$,} depending on the choice of NME, the predicted 
half-life in $^{130}\rm{Te}$, $^{100}\rm{Mo}$, $^{82}\rm{Se}$
 can vary over a wide range and may be    constrained 
by CUORICINO \cite{cuo} and NEMO 3 \cite{nemo3}. However, if we consider smaller $m^{\nu}_{\mathrm{ee}}$, more sensitive experiments are needed 
and the searches for $(\beta \beta)_{0\nu}$-decay will be even more challenging than in the case of light 
neutrino mass only.

The existence of heavy and/or light sterile neutrinos can be easily implemented in 
 seesaw scenarios, such as  Type-I, Extended or Type-I+Type-II seesaw,  in which  
the cancellation between light and heavy neutrino exchanges can be realized. 
In these models   light neutrino masses can be generated either at tree  or   loop level. 
An Extended seesaw scenario allows for sterile neutrino in the 100 MeV mass range while having 
sufficiently large mixing angles with electron neutrinos 
and a cancellation between light active and heavy sterile neutrino contribution. The 
case in which both light and heavy sterile neutrinos are at play can be realized in a further 
extension of the model above in which a light neutrino mass come from a Type-II seesaw framework. 
In all of the cases,  very precise values of masses and mixings are  needed  
to induce a  cancellation and  require a high level of  fine-tuning.

\vspace*{0.4cm}

{\bf{  Acknowledgements} } 

\vspace*{0.3cm}
The authors  acknowledge the partial support of the ITN INVISIBLES (Marie Curie Actions, PITN-GA-2011-289442). 
SP thanks SISSA for hospitality,  where part of this study has been conducted.  MM thanks ICTP for hospitality. 
The authors thank S.~ Petcov  for  discussions in the initial stage of this work.

\newpage

\vspace*{0.4cm}
 \begin{center}{\bf{Appendix}}\end{center}
\vspace*{0.2cm}

We consider $n_l$ generations of light sterile neutrinos of masses $m_{4_k}$ ($k=1,2,..n_l$) and $n_h$ generations of heavy sterile neutrinos of masses $M_{N_j}$ ($j=1,2,...n_h$). The active-light and active-heavy sterile mixings are $U_{\mathrm{e}4_k}$ and $V_{\mathrm{e}N_j}$,  respectively. 
The half-life of neutrinoless double beta decay is 
\begin{equation}
\frac{1}{T^{0\nu}_{1/2}}=G_{0\nu}\left|\mathcal{M}_{\nu}\eta_{\nu}+\mathcal{M}_N\eta_N\right|^2, 
\label{eq1}
\end{equation}
where $G_{0\nu}$ is the phase space factor, $\mathcal{M}_{\nu}$ and $\mathcal{M}_N$ are the nuclear matrix elements
corresponding to the light and heavy neutrino exchange. In the limit when the mass of the sterile neutrinos are far
from the intermediate momentum exchange, i.e. 
$m^2_{4_k}<<|p^2|\, \rm{MeV}^2$ and $M^2_{N_j}>>|p^2|\, \rm{MeV}^2$,  the factors $\eta_{\nu}$ and $\eta_N$ are 
$\eta_{\nu}=\frac{\Sigma_i m_i U^2_{\mathrm{e}i}+\Sigma_k m_{4_k}U^2_{\mathrm{e}4_k}}{m_\mathrm{e}}$ and $\eta_N=\sum_j \frac{V^2_{\mathrm{e}N_j}m_\mathrm{p}}{M_{N_j}}$, 
where we include the contributions from light active, light sterile as well as heavy sterile neutrinos.
An equivalent way of description is 
 \begin{equation}
\frac{1}{T^{0\nu}_{1/2}}=K_{0\nu}\left|\frac{ m_i U^2_{\mathrm{e}i}+m_{4_k}U^2_{\mathrm{e}4_k}}{p^2}-\frac{V^2_{\mathrm{e}N_j}}{M_{N_j}}\right|^2.
\label{eq5}
\end{equation}
In the above,  the indices $i,j,k$ are summed over,  $p^2=-m_\mathrm{e}m_\mathrm{p} \frac{\mathcal{M}_{N}}{\mathcal{M}_{\nu}}$ and 
and $K_{0 \nu}=G_{0\nu}(m_\mathrm{p}\mathcal{M}_N)^2$. The generic expression
that is valid even for the mass range $m^2_{4_k} \simeq |p^2|\, \rm{MeV}^2$ and $M^2_{N_j} \simeq |p^2|\, \rm{MeV}^2$ is 
the following:
\begin{equation}
\frac{1}{T^{0\nu}_{1/2}}=K_{0 \nu}\big|\mathcal{\theta}^2\frac{m}{p^2-m^2}\big|^2,
\end{equation}
where $\mathcal{\theta}$ is the mixing angle and $m$ is the mass of corresponding neutrino state. Following this, 
we have the generic expression,
\begin{equation}
\frac{1}{T^{0\nu}_{1/2}}=K_{0 \nu}\left|\frac{U^2_{\mathrm{e}i}m_i}{p^2}+\frac{U^2_{\mathrm{e}4_k}m_{4_k}}{p^2-m^2_{4_k}}+\frac{V^2_{\mathrm{e}N_j}M_{N_j}}{p^2-M^2_{N_j}}\right|^2.
\label{eqnexact}
\end{equation}
 In the limit that light and heavy sterile neutrinos  have masses far from momentum exchange scale, one will obtain  
Eq.~\ref{eq5}. For concreteness,  we consider the case of one light sterile neutrino and one heavy sterile neutrino with masses $m_4$ and $M_N$, respectively. 
If the light and heavy neutrino contributions cancel each other in   isotope A, then the half-life $T^{0\nu}_{1/2}(\rm{A})$ is infinite and we have
\begin{equation}
\frac{|V^2_{\mathrm{e}N} M_N|}{|p^2_{\rm{A}}|+M^2_N}= \left|\frac{U^2_{\mathrm{e}i}m_i}{|p^2_{\rm{A}}|}+\frac{U^2_{\mathrm{e}4}m_{4}}{|p^2_{\rm{A}}|+m^2_{4}}\right|,
\label{eqc}
\end{equation}
  The expression simplifies considerably, if we neglect the three light active neutrino contribution. In addition, If
light sterile and heavy sterile neutrino contribution saturate the bound or claimed value of half-life $T^{0\nu}_{1/2}({B})$ of any other isotope B, then  the  
contour of active-light sterile neutrino mixing is   
\begin{equation}
|U^2_{\mathrm{e}4}|^2=\frac{m^{-2}_4}{K^{{B}}_{0\nu} T^{0\nu}_{1/2}\big({B})}\frac{1}{{(\frac{1}{|p^2_{\rm{B}}|+m^2_4}-\frac{1}{|p^2_{A}|+m^2_4}
\frac{|p^2_{\rm{A}}|+M^2_N}{|p^2_{\rm{B}}|+M^2_N}\big)^2}}.
%-2(\frac{|p^2_{\rm{A}}|+M^2_N}{|p^2_{\rm{B}}|+M^2_N})\frac{1}{(|p^2_{A}|+m^2_4)(|p^2_{B}|+m^2_4)}})}
\end{equation}
Using Eq.~\ref{eqc}, the contours for 
active-heavy sterile neutrino mixing can be obtained:
%would be 
\begin{equation}
|V^2_{\mathrm{e}N}|^2=\frac{M^{-2}_N}{K^{{B}}_{0\nu}T^{0\nu}_{1/2}({B})}\frac{1}{{\big(\frac{1}{|p^2_{\rm{B}}|+M^2_N}-\frac{1}{|p^2_{A}|+M^2_N}
\frac{|p^2_{\rm{A}}|+m^2_4}{|p^2_{\rm{B}}|+m^2_4}\big)^2}}.
%-2(\frac{|p^2_{\rm{A}}|+m^2_4}{|p^2_{\rm{B}}|+m^2_4})\frac{1}{(|p^2_{A}|+M^2_N)(|p^2_{A}|+M^2_N)}})}
\end{equation}
%This equations can be applied, for {\it e.g.,} when cancellation is effective in $^{136}\rm{Xe}$
These generic equations can be applied for e.g. $^{136}\rm{Xe}$, $^{76}\rm{Ge}$, or for any other isotopes.

\end{document}